\documentclass{aa}  

\usepackage{graphicx}
\usepackage{txfonts}
\usepackage{hyperref}
\usepackage[version=4]{mhchem}

\usepackage{listings}
\usepackage{siunitx}
\usepackage{tabularx}
\usepackage{subcaption}
\usepackage{threeparttable}

\usepackage{natbib}
\bibpunct{(}{)}{;}{a}{}{,} 

\DeclareSIUnit\years{years}
\DeclareSIUnit\pc{pc}
\DeclareSIUnit\gauss{G}

\begin{document} 

   \title{Tracing shock type with chemical diagnostics}

   \subtitle{An application to L1157}

   \author{T. A. James
          \inst{1}
          \and
          S. Viti\inst{1}
          \and
          J. Holdship\inst{1}
          \and
          I. Jimenez-Serra\inst{2}
          }

   \institute{Department of Physics and Astronomy, 
              University College London, 
              Gower Street, 
              London WC1E 6BT,
              UK \\
              \email{tjames@star.ucl.ac.uk}
         \and
             Centro de Astrobiologia (CSIC, INTA), 
             Ctra. de Ajalvir, km. 4, 
             Torrejón de Ardoz, 
             28850 Madrid, 
             Spain
             }

   \date{Received; accepted}

 
  \abstract
   {}
   {The physical structure of a shock wave may take a form unique to its shock type, implying that the chemistry of each shock type is unique as well. We aim to investigate the different chemistries of J-type and C-type shocks in order to identify unique molecular tracers of both shock types. We apply these diagnostics to the protostellar outflow L1157 to establish whether the B2 clump could host shocks exhibiting type-specific behaviour. Of particular interest is the L1157-B2 clump, which has been shown to exhibit bright emission in \ce{S}-bearing species and \ce{HNCO}.}
   {We simulate, using a parameterised approach, a planar, steady-state J-type shock wave using \lstinline{UCLCHEM}. We compute a grid of models using both C-type and J-type shock models to determine the chemical abundance of shock-tracing species as a function of distance through the shock and apply it to the L1157 outflow. We focus on known shock-tracing molecules such as \ce{H_{2}O}, \ce{HCN}, and \ce{CH_{3}OH}.}
   {We find that a range of molecules including \ce{H_{2}O} and \ce{HCN} have unique behaviour specific to a J-type shock, but that such differences in behaviour are only evident at low $v_{s}$ and low $n_{H}$. We find that \ce{CH_{3}OH} is enhanced by shocks and is a reliable probe of the pre-shock gas density. However, we find no difference between its gas-phase abundance in C-type and J-type shocks. Finally, from our application to L1157, we find that the fractional abundances within the B2 region are consistent with both C-type and J-type shock emission.}
  {}

   \keywords{astrochemistry -- ISM: evolution -- ISM: individual objects (L1157) -- ISM: molecules -- stars: protostars
               }

   \maketitle
%
%
\section{Introduction}
    Astrophysical shocks represent prominent catalysts for chemical evolution in the Interstellar Medium (ISM). The low signal-speed within the ambient ISM leads to a variety of different astrophysical events driving supersonic flows that form shocks, from cloud-cloud collisions \citep[e.g.][]{cloudCloud} to bipolar outflows emanating from protostellar objects \citep[e.g.][]{outflowFirst,shuOutflows,outflowShock}. The different ambient gas conditions that a supersonic flow can be driven into leads to the production of different shock types. \citet{draineShocks} initially defined two shock types, C (continuous) type shock and J (jump) type shock, with subsequent computational work by \citet{cjShocksChieze} and \citet{cjShocksFlower} defining a third, CJ (mixed) type shock. 

    Unlike C-type shocks, which typically arise in regions with a magnetic field and low degree of fractional-ionisation, J-type shocks arise in regions whereby only a negligible magnetic field is present \citep{draineShocks}. The negligible magnetic field within a J-type shock has further consequences in that it does not act to limit the compression through the shock, thus allowing a higher peak temperature to be reached within the shock-front (relative to a C-type shock). Owing to this, J-type shocks are thought to exhibit far more destructive chemistry than a C-type shock counterpart. An analytic description of a C-type shock therefore requires equations of MHD and multiple fluid components, whilst J-type shocks can be described by hydrodynamics equations and a single fluid alone. 
    
    Typically, such descriptions are implemented in MHD codes such as \lstinline{mhd_vode} \citep{mhd_vode}. However, such approaches to modelling incur a large amount of computational expense, necessitating compromises in the complexity and size of the chemical network used. By using a parameterised form of the physical structure of the shock, as \citet{cshockParameter} did with their C-type shock parameterisation, it is possible to preserve an approximation of the shock structure whilst significantly reducing computational complexity, thus allowing the computation of far more complex chemistry.
    
    This is particularly important owing to the complex chemistry that is influenced by shocks. In particular, shocks can drive chemical reactions that would otherwise be highly unlikely to occur under quiescent ISM conditions. For example, the reaction \ce{O + H_{2} -> OH + H} has an activation barrier of $\approx 1$ \si{\electronvolt} and would therefore require temperatures $> 1000$ \si{\kelvin}, which are easily achievable within shocks, to initiate \citep{baulchChem,vanDischoeckReview,williamsVitiBook}. 
    
    It is through such reactions that the axiom of unique chemistry as a diagnostic of prior physical events is drawn. Further reinforcing this axiom is interstellar chemistry's high density and temperature dependence, thus rendering the composition of the ISM highly sensitive to dynamical environmental effects. Shocks are ubiquitous sources of such change within the ISM, and therefore represent prominent sources of chemical enrichment in early star-forming environments.
    
    Observations of shocked regions allow effective probes of the shock chemistry. Recent high-resolution spectroscopy programmes such as \lstinline{ASAI} \citep{asaiSurvey}, \lstinline{CHESS} \citep{chessSurvey} and \lstinline{WISH} \citep{wishSurvey} permit unprecedented insight into not just early stages of star formation, but also the violent events that initially drive shocks into these regions. The bipolar outflow in L1157 \citep{L1157Disco} is an example of a prototypical protostellar outflow observed during these programmes. Observations of outflows cannot, however, provide insight into either the physical or subsequent chemical evolution of the shock through time, instead only capturing a static snapshot of the conditions. Modelling shock-induced chemistry is therefore one of the only methods of following the evolution of an inherently time-dependent chemical process in astrophysics.

    The role that dust grains play in interstellar chemistry is also of paramount importance.  Molecules in the gas-phase may freeze on to the surface of dust grains, thereby depleting their gas-phase abundance by changing state. Processes such as successive hydrogenation on dust grains are thought to be the mechanism responsible for such complex organic molecule formation as \ce{CH_{3}OH} \citep{comIces,Fuchs2009AstrophysicsApproach}. Importantly this method also presents a viable solution to the cold gas-phase abundance problem whereby molecules are observed in the gas phase at temperatures well below their gas-phase formation temperature. Under the influence of a sputtering, grain-grain collision or desorption event (thermal or non-thermal), the molecule may be released from the surface of the dust-grain directly into the gas phase. This complex interplay between the gas-phase and dust-grain chemistry essentially chemically couples the two phases. It is therefore vitally important when modelling interstellar chemistry that both gas-phase and dust-grain reactions included within the reaction network are accurate and comprehensive for the relevant molecules. 
    
    In practice, the only way one can hope to distinguish between the two types of shock is to systematically determine the effects of each shock type and hence compare the resultant chemical distinctions. Our goal in this paper is to identify molecular tracers of a J-type shock by using such a technique and apply it to a shocked region of L1157 thought to be exhibiting signatures of both C-type and J-type shock behaviour. We therefore make extensive use of the C-type shock module, based on \citet{cshockParameter}, that is already implemented in \lstinline{UCLCHEM} \citep{UCLCHEM}. To that goal, we present in Section \ref{sec:L1157} an overview of L1157. We present in Section \ref{sec:jShockModel} a parameterised model of a J-type shock built for the astrochemical code \lstinline{UCLCHEM}. In conjunction with the pre-existing C-type shock model based upon \citet{cshockParameter} we investigate in Section \ref{sec:results} the chemical distinctions between J-type and C-type shocks to identify unique chemical tracers of both shock types. Section \ref{L1157Apply} applies these results by comparing them to enhanced abundances with shocked regions of the L1157 outflow.
    
    \section{L1157} \label{sec:L1157}
    
    \begin{table*}[t]
        \centering
        \caption{Abundances $\chi$ of known shock enhanced molecules and their enhancement factors $f$ (relative to $\chi(0)$) in the two L1157 knots B1 and B2. $\chi(0)$ is the fractional abundance of each molecule measured towards the central driving protostar L1157-mm.}
        \begin{tabular}{l l l l l l l}
            \hline\hline                        
            Molecule & $\chi(0)$ & $\chi(B1)$ & $\chi(B2)$ & $f(B1)$ & $f(B2)$ & Reference \\    
            \hline
            \ce{CH_{3}OH} & $4.5\times10^{-8}$ & $0.4-1.9\times10^{-5}$ & $2.2\times10^{-5}$ & $300-400$ & $500$ & (1) \\
            \ce{HCN} & $3.6\times10^{-9}$ & $3.3\times10^{-7}$ & $5.5\times10^{-7}$ & $90$ & $150$ & (1) \\ 
            \ce{SO} & $\sim 5.0\times10^{-9}$ & $2.0-3.0\times10^{-7}$ & $2.0-5.0\times10^{-7}$ & $50-70$ & $60-100$ & (1) \\ 
            \ce{SO_{2}} & $\lessapprox 3.0\times10^{-8}$ & $2.1\times10^{-7}$ & $5.7\times10^{-7}$ & $\sim 8$ & $\sim 20$ & (1) \\ 
            \ce{H_{2}O} & (...) & $1 \times 10^{-4}$ & $1 \times 10^{-6}$ & (...) & (...) & (2) \\
            \ce{HNCO} & $0.3-1.2 \times 10^{-9}$ & $4.3-17.9 \times 10^{-9}$ & $25-96 \times 10^{-9}$ & $\sim 15$ & $\sim 80$ & (3) \\
            \hline
        \end{tabular}
        \begin{tablenotes}
            \small
            \item References: (1) \citet{B2Bright}; (2) \citet{Vasta2012}; (3) \citet{B2HNCO}
        \end{tablenotes}
        \label{tab:enhance}
    \end{table*}
    
    At $250$ \si{\pc} \citep{L1157d}, L1157 is a nearby region that comprises a central class-0 protostar, L1157-mm, that in turn drives a bipolar outflow. The observed outflow produces a red-shifted lobe to the North and a blue-shifted lobe to the South that are aligned with the protostar's rotation axis. A degree of symmetry is observed in these lobes, however the geometry of lobe sub-structure indicates the presence of an underlying precessing jet \citep{B2JShock}. This precession allows periodic ejection events to create complex structures enhanced by shocks \citep{L1157}. The Southern lobe hosts two intriguing examples of such shock events: the clumps B1 and B2, which are themselves located within larger cavities C1 and C2. As a result, L1157 is considered to be one of the best laboratories for astrochemistry \citep{L1157Disco,L1157Prototype}.
    
\begin{figure} 
        \centering
        \includegraphics[scale=.6]{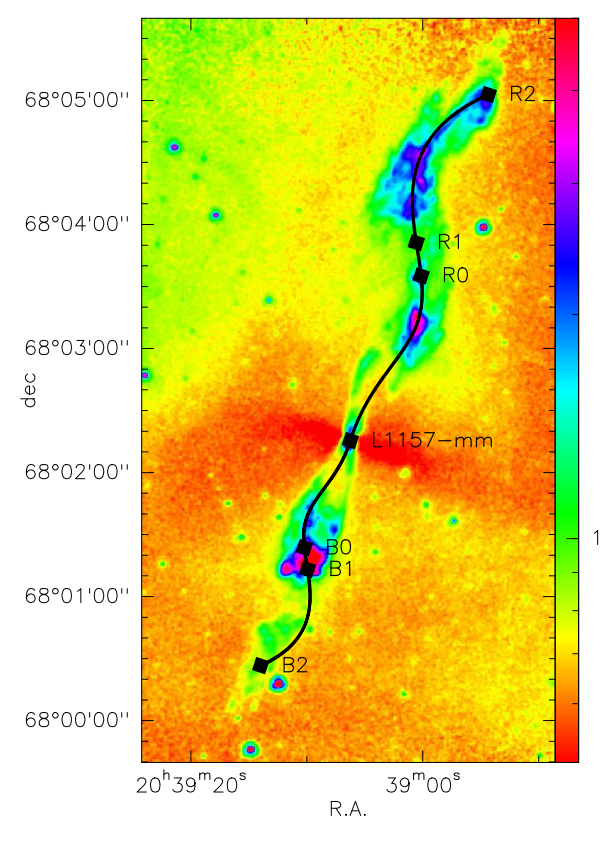} 
        \caption{Spitzer/IRAC \num{8} \si{\micro\meter} image of the L1157 outflow. \citep{L1157Precession}. Shown as black squares are the shock events B0, B1 and B2. The class-0 protostar L1157-mm that drives the outflow is also labelled. The black line overplotted is the precession model thought to be responsible for the creation of the observed knots. As is visible here, B2 is far less intense in emission than B0/B1.} \label{fig:L1157}
    \end{figure}

    Figure \ref{fig:L1157} shows Spitzer/IRAC \num{8} \si{\micro\meter} observations by \citet{L1157Precession}. Labelled are the knots B0, B1 and B2 alongside the central driving protostar L1157-mm and the proposed precession model from \citet{L1157Precession}.
  
    It has since been found that B1 and B2 themselves host low-velocity clumps. \cite{milenaB2}, using PdB interferometric observations, showed that nine clumps exist within the B1 and B2 structure, thus giving rise to even further complexity within the Southern lobe. This substructure is thought to arise from L1157-mm's precession, which creates complex knots driven by shock-activity produced by the host outflow.
    
    \subsection{L1157-B1} \label{sec:B1}
    B1 is the brightest clump within the L1157 region and thus the subject of significant study. It is warm and young, exhibiting kinetic temperatures between $T \approx 80-100$ K and age $t \approx 1000$ years. In comparison B2 is colder and older with $T \approx 20-60$ K and $t \approx 4000$ years \citep{B1Phys,L1157}. \cite{viti2011} first showed, with confirmation by \cite{milenaB1}, that B1 is likely produced by a non-dissociative, C-type shock with pre-shock density $n_{H} \geq 10^{4}$ \si{\per\centi\meter\cubed} and $v_{s} \approx 40$ \si{\kilo\meter\per\second}, leading to a maximum obtainable temperature of $\sim 4000$ \si{\kelvin}. 
    
    \subsection{L1157-B2} \label{sec:B2}
    Being less intense in most emission lines, B2 has been subject to far less study. B2 is, however, brighter than B1 in most sulphur-bearing species as well as \ce{HNCO} \citep{B1Phys,B2Bright,B2HNCO}. \citet{B1Phys} specifically finds that \ce{SO} and \ce{SO_{2}} exhibit enhancement factors within L1157-B2 (relative to L1157-mm) of between $60 - 100$ and $20$, respectively. Meanwhile, they also find that the enhancement factors for L1157-B1 are $50 - 70$ and $8$.  \ce{HNCO} is thought to form efficiently on grain surfaces, whilst \ce{S}-bearing species like \ce{SO} and \ce{SO_{2}} form in the gas-phase with sputtered \ce{S} from the grains themselves \citep{allenHNCO,charnleySulphur,garrodHNCO}. The older dynamical age of B2 relative to B1 could lend credence to the idea that B2 has simply had more time than B1 to chemically process the sputtered material, hence the more luminous species like \ce{HNCO} and \ce{S}-bearing species. Table \ref{tab:enhance} lists further molecules observed within L1157 and their enhancement factors, where $f_{enhance} = \chi(R)/\chi(0)$. Importantly these enhancement factors, as well as their associated abundances, are subject to large uncertainties arising from the assumption that the observed lines are both optically thin and thermalised.

    To date studies such as those by \cite{B2JShock} have not yet been able to determine with certainty the prevalent shock type within B2, though \citeauthor{B2JShock} does allude to the possibility of a J-type shock component within L1157-B2. \citet{L1157Profiles} use \ce{NH_{3}} and \ce{H_{2}O} abundances, alongside model predictions, to trace shock temperature within L1157's lobes. \citeauthor{L1157Profiles} finds that whilst a proper line radiative transfer model is needed for proper computation, the best matching model for L1157-B2 is one with $n_{H} \approx 10^{3}$ \si{\per\centi\meter\cubed} and $v_{s} \approx 10$ \si{\kilo\meter\per\second}.

%
%

\section{Shock modelling} \label{sec:jShockModel}
    
    Our parameterised model is based on the MHD code \lstinline{mhd_vode} \citep{mhd_vode}. \lstinline{mhd_vode} is an ideal-MHD, 1D, two-fluid simulation of both C-type and J-type shocks that computes chemistry in parallel with its physics. This model is built as a module to the time-dependent chemical code \lstinline{UCLCHEM} \citep{UCLCHEM}. 
    
    \lstinline{UCLCHEM} is a diverse code, and its modularised functionality lends itself to a host of different astrochemical problems and environments. For a full description of \lstinline{UCLCHEM}'s operation see \citet{UCLCHEM} as well as the documentation hosted at \href{https://uclchem.github.io/}{https://uclchem.github.io/}. In brief, \lstinline{UCLCHEM} is constructed so as to follow a two-phase computation. Firstly an ambient medium of user-supplied temperature, density and chemical composition undergoes an isothermal collapse as described by \citet{rawlingsCollapse} to a user-supplied final density. The chemical composition of a 1D parcel is therefore followed during collapse, and thus informs the chemical conditions for phase 2. During phase 2, the relevant physics supplied via a user module is computed and used to inform the rates of reactions within the chemical network. Our J-type shock module is built so as to follow this methodology. 
    
    \subsection{J-type shock parameterisation}
    To construct our parameterised model we first noted that, as described by \citet{zeldovich}, shocks can generally be discretised into four regions: the precursor, the shock-front, the post-shock relaxation layer and the thermalisation layer. We neglect the radiative precursor component of the shock in our models, as J-type shocks with $v_{s} < 80$ \si{\kilo\meter\per\second} have been found to have negligible radiative precursor components, therefore playing no role in either the shock structure or the shock chemistry \citep{radiativePrecursor,precursorJustification}. We also neglect the thermalisation layer, instead focusing on the shock-front and the post-shock relaxation layer as sole sources of chemical evolution. We assume that the post-shock gas cools to its initial temperature in the post-shock relaxation layer.
    
    To build the shock-front, we ran a grid of \lstinline{mhd_vode} models with the magnetic field $B=0$ \si{\gauss} and interstellar values for cosmic-ray ionisation rate $\zeta_{CR}$ and radiation field, so as to quantify the trend in temperature and density, as well as the shock-front duration $t_{front}$, across the parameter space we were exploring. $t_{front}$, in units of \si{\second}, is described by Equation \ref{eq:t-front}.
    
    \begin{equation} \label{eq:t-front}
        t_{front} = \frac{\left( \sqrt{2} \pi \left(5.76 \times 10^{-16}\right)^{-1}\right)}{v_{s} \times 10^{6}}
    \end{equation}
    
    Where $v_{s}$ represents the initial shock velocity in \si{\kilo\meter\per\second}. The increase in temperature and density within the shock-front was found to be best described by $T = T_{max}\left(t/t_{front}\right)^{2}$ \label{eq:shock-front-T} in \si{\kelvin} and $n_{H} = 4n_{H_initial}\left(t/t_{front}\right)^{4}$ \label{eq:shock-front-n} in \si{\per\centi\meter\cubed}. For $t < t_{front}$ we assume that the Rankine-Hugoniot conditions \citep{rankine,hugoniot} hold such that the density $n_{H}$ increases to $\approx 4$ times its initial value whilst the temperature $T$ increases to its maximum obtainable value, $T_{max}$. $T_{max}$ is determined by $T_{max} = 5\times10^{3} \left(v_{s}/10\right)^{2}$ in \si{\kelvin} \citep{williamsVitiBook}.
    
    After the shock-front, the shocked gas begins to cool, representing the post-shock relaxation layer where $t > t_{front}$ and $t < t_{shock}$. $t_{shock}$ was obtained by fitting a polynomial to a range of shock timescales from \lstinline{mhd_vode} models and is described by Equation \ref{eq:t-shock}.
    
    \begin{equation} \label{eq:t-shock}
        t_{shock} = \frac{t_{year}\times10^{6}}{n_{H_{initial}}}
    \end{equation}
    
    Where $t_{year}$ is the number of seconds in $1$ year and $n_{H_{initial}}$ is the initial pre-shock number density in \si{\per\centi\meter\cubed}. The factor of $10^{6}$ acts as a normalising density such that $t_{shock}$ has units of \si{\second}.
    
    Within this layer, the temperature and density equations take the forms described in Equations \ref{eq:post-shock-T} and \ref{eq:post-shock-n}.
    
    \begin{equation} \label{eq:post-shock-T}
        T = T_{max}e^{-\lambda_{T}\left(\frac{t}{t_{shock}}\right)}
    \end{equation}
    
    \begin{equation} \label{eq:post-shock-n}
        n = 4n_{initial}e^{\lambda_{n}\left(\frac{t}{t_{shock}}\right)}
    \end{equation}
    
    Equation \ref{eq:post-shock-T} has units of \si{\kelvin}, whilst Equation \ref{eq:post-shock-n} has units of \si{\per\centi\meter\cubed}. This therefore allows the gas to cool following a decaying exponential law, whilst the gas also increases in density to $n_{H_{max}}$, which is itself derived from \lstinline{mhd_vode} grids. $n_{H_{max}}$ is defined as $n_{H_{max}} = \left(v_{s} \times n_{H_{initial}}\right)\times10^{2}$ in units of \si{\per\centi\meter\cubed}. The constants $\lambda_{T}$ and $\lambda_{n}$ in Equations \ref{eq:post-shock-T} and \ref{eq:post-shock-n} are described by $\lambda_{T} = ln \left( \frac{T_{max}}{T_{initial}} \right)$ and $\lambda_{n} = ln \left( \frac{n_{max}}{n_{initial}} \right)$. At $t > t_{shock}$, we assume that the gas has cooled back to its initial temperature $T_{initial}$. We assume a steady-state profile for both $T$ and $n$, and discuss the validity of this approximation in Section \ref{sec:modelComparison}.
    
    \subsection{C-type shock parameterisation} \label{sec:cShock}
    \lstinline{UCLCHEM} implements a version of the parameterised C-type shock from \citet{cshockParameter}. The \lstinline{UCLCHEM} implementation is described in more detail, as well as demonstrated to good effect, in \citet{UCLCHEM}. 
    
    Similarly to the J-type shock parameterisation presented in Section \ref{sec:jShockModel}, \citet{cshockParameter} approximates the physical shock structure using analytical equations for $T$ and $n_{H}$ alongside the velocity of the ions and neutrals, $v_{i}$ and $v_{n}$ respectively (see Appendix A of \citet{cshockParameter} for further details). They also make use of results from \citet{draineRobergeDalgarno} to parameterise the maximum shock temperature $T_{max}$ as a function of shock velocity $v_{s}$. It is this temperature that is shown for the C-type shock in Table \ref{tab:grid}.
    
    \citet{cshockParameter} also present, in Appendix B, a fractional sputtering treatment of grain mantle species such \ce{Si}, \ce{CH_{3}OH}, and \ce{H_{2}O}. \lstinline{UCLCHEM} now supports this sputtering implementation. In summary, rather than an instantaneous ejection of the mantle into the gas phase when the saturation time $t_{sat}$\footnote{$t_{sat}$ is defined as the time for which the logarithmic difference of the \ce{Si} abundance between two consecutive timesteps $t_{i+1}$ and $t_{i}$ is $\left| \log_{10} \chi(m_{i+1}) - \log_{10} \chi(m_{i}) \right| < 0.1$.} is exceeded, only a fraction of the species abundance will be released from the mantles and/or ices at any given timestep providing the drift velocity between the neutrals and ions, as well as the impact energy, is sufficient to sputter material. 
    
    Of critical importance in C-type shock formation is the magnetic field, $B$. \lstinline{UCLCHEM}'s C-type shock implementation assumes the $B$-field (in \si{\micro\gauss)} scales according to the emperical law defined in \citet{draineRobergeDalgarno}, i.e. $B_{0}=b_{0}\sqrt{n_{H}}$ where $b_{0}$ is the magnetic scaling parameter and $n_{H}$ the Hydrogen number density. Much like \citet{draineRobergeDalgarno}, we fix $b_{0}$ as $1$, thus allowing the magnetic field to scale with $\sqrt{n_{H}}$ as defined in Table $4$ of \citet{draineRobergeDalgarno}. According to this relation, at $n_{H}=10^{3}$ \si{\per\centi\meter\cubed} the magnetic field has a field strength of $B_{0}=10$ \si{\micro\gauss} whilst at at $n_{H}=10^{6}$ \si{\per\centi\meter\cubed} the magnetic field has field strength $B_{0}=1$ \si{\milli\gauss}, both of which are consistent with Table 4 of \citet{draineRobergeDalgarno}.
    
    \subsection{Computational grid}
    \citet{L1157Profiles} finds the best fit profile to \ce{NH_{3}} and \ce{H_{2}O} abundances in L1157-B2 is one with $v_{s}=10$ \si{\kilo\meter\per\second} and $n_{H}=10^{3}$ \si{\per\centi\meter\cubed}, and we use this as to inform our choice of initial conditions for our grid of models. 
    
    Table \ref{tab:grid} shows the range of parameters used to compute this grid. For a J-type shock $T_{max}$ is determined as discussed, whilst for a C-type shock $T_{max}$ is determined according to the parameterisation discussed in \citet{cshockParameter} (see Section \ref{sec:cShock}).
    
    \begin{table}
        \caption{Grid of models used to compute simulations. The velocity $v_{s}$, density $n_{H}$ and maximum temperature achieved in both C-type and J-type shocks, $T_{max}$, is shown. Each model is run twice: once for a C-type shock and once for a J-type shock.}              
        \label{tab:grid}      
        \centering                                      
        \begin{tabular}{c c c c c}          
        \hline\hline                        
        Model & $n_{H}$ [\si{\per\centi\meter\cubed}] & $v_{s}$ [\si{\kilo\meter\per\second}] & \multicolumn{2}{c}{$T_{max}$ [\si{\kelvin}]} \\
          &                     &                       & C-type           & J-type \\
        \hline                                  
            1 & $10^{3}$ & $5$ & $85$ & $1250$  \\      
            2 & $10^{4}$ & $5$ & $85$ & $1250$ \\
            3 & $10^{5}$ & $5$ & $85$ & $1250$ \\
            4 & $10^{6}$ & $5$ & $85$ & $1250$ \\
            5 & $10^{3}$ & $6$ & $131$ & $1800$ \\
            6 & $10^{4}$ & $6$ & $131$ & $1800$ \\
            7 & $10^{5}$ & $6$ & $131$ & $1800$ \\
            8 & $10^{6}$ & $6$ & $131$ & $1800$ \\
            9 & $10^{3}$ & $7$ & $178$ & $2450$ \\
          10 & $10^{4}$ & $7$ & $178$ & $2450$ \\
          11 & $10^{5}$ & $7$ & $178$ & $2450$ \\
          12 & $10^{6}$ & $7$ & $178$ & $2450$ \\
          13 & $10^{3}$ & $8$ & $225$ & $3200$ \\
          14 & $10^{4}$ & $8$ & $225$ & $3200$ \\
          15 & $10^{5}$ & $8$ & $225$ & $3200$ \\
          16 & $10^{6}$ & $8$ & $225$ & $3200$ \\
          17 & $10^{3}$ & $9$ & $273$ & $4050$ \\
          18 & $10^{4}$ & $9$ & $273$ & $4050$ \\
          19 & $10^{5}$ & $9$ & $273$ & $4050$ \\
          20 & $10^{6}$ & $9$ & $273$ & $4050$ \\
          21 & $10^{3}$ & $10$ & $323$ & $5000$ \\
          22 & $10^{4}$ & $10$ & $323$ & $5000$ \\
          23 & $10^{5}$ & $10$ & $323$ & $5000$ \\
          24 & $10^{6}$ & $10$ & $323$ & $5000$ \\
          25 & $10^{3}$ & $11$ & $373$ & $6050$ \\
          26 & $10^{4}$ & $11$ & $373$ & $6050$ \\
          27 & $10^{5}$ & $11$ & $373$ & $6050$ \\
          28 & $10^{6}$ & $11$ & $373$ & $6050$ \\
          29 & $10^{3}$ & $12$ & $424$ & $7200$ \\
          30 & $10^{4}$ & $12$ & $424$ & $7200$ \\
          31 & $10^{5}$ & $12$ & $424$ & $7200$ \\
          32 & $10^{6}$ & $12$ & $424$ & $7200$ \\
          33 & $10^{3}$ & $13$ & $477$ & $8450$ \\
          34 & $10^{4}$ & $13$ & $477$ & $8450$ \\
          35 & $10^{5}$ & $13$ & $477$ & $8450$ \\
          36 & $10^{6}$ & $13$ & $477$ & $8450$ \\
          37 & $10^{3}$ & $14$ & $530$ & $9800$ \\
          38 & $10^{4}$ & $14$ & $530$ & $9800$ \\
          39 & $10^{5}$ & $14$ & $530$ & $9800$ \\
          40 & $10^{6}$ & $14$ & $530$ & $9800$ \\
          41 & $10^{3}$ & $15$ & $585$ & $11250$ \\
          42 & $10^{4}$ & $15$ & $585$ & $11250$ \\
          43 & $10^{5}$ & $15$ & $585$ & $11250$ \\
          44 & $10^{6}$ & $15$ & $585$ & $11250$ \\
        \hline                                             
        \end{tabular}
    \end{table}
    
    We also account for the initial C-type shock conditions published by other authors so as to verify the feasibility of C-type shock formation at the conditions considered. For example \citet{UCLCHEM} identifies C-type shock-tracing molecules for a range of different physical shock conditions to a lower limit of $v_{s}=10$ \si{\kilo\meter\per\second} and $n_{H}=10^{3}$ \si{\per\centi\meter\cubed}. Furthermore, \citet{draineRobergeDalgarno} identify the maximum shock temperature for a range of different C-type shocks with a lower limit of $v_{s}=5$ \si{\kilo\meter\per\second} and $n_{H}=10^{2}$ \si{\per\centi\meter\cubed} with a $B$ field defined by $B=10$ \si{\micro\gauss}. Finally, \citet{godardShocks} investigate the formation of a range of different shock types under different $B$ fields and irradiated conditions. They highlight C-type shocks forming between $v_{s}=5-20$ \si{\kilo\meter\per\second} and $n_{H}=10^{2}-10^{5}$ \si{\per\centi\meter\cubed} under a range of $B$ fields from $B=1$ \si{\micro\gauss} to $B=3$ \si{\milli\gauss}. Our parameters fit comfortably into this published range and we therefore assume that C-type shock formation at these conditions is entirely feasible.
    
    For each $v_{s}$ and $n_{H}$ within Table \ref{tab:grid}, the fractional abundance of 215 individual molecules, including \ce{H_{2}O}, \ce{HCN}, \ce{CH_{3}OH}, \ce{SO} and \ce{SO_{2}}, was computed for both C-type and J-type shocks. This was achieved by coupling the physical shock computations from within the physics modules of \lstinline{UCLCHEM} to a chemical network of 2456 reactions. Further details of the network are discussed later in this section. We plot the fractional abundance of a molecule against distance through the shock, up to the C-type shock dissipation length as determined by \citet{cshockParameter}. The dissipation length is defined as the distance over which the velocity of the ions and neutrals equalises \citep{draineShocks}. As a J-type shock consists of one fluid that encompasses both ions and neutrals, the concept of a dissipation length does not apply. Instead, we plot the J-type shock fractional abundance up to the cooling length of the shock, beyond which the gas has reached equilibrium. As the fluids within a C-type shock also reach equilibrium at the dissipation length, we assume the two distance scales are comparable.
    
    Using these plots, the abundance trends were then compared between shock types to better understand the behaviour of species under different shock conditions. Of particular interest in this study was the enhancement factors observed in Table \ref{tab:enhance}, as this forms the signature of shock passage and therefore the best diagnostic of shock type in a shocked region.
    
    Principal to this enhancement factor analysis is the assumption that the pre-shock gas is homogeneous throughout L1157 and the surrounding region, therefore allowing the fractional abundance at $t \approx 0$ \si{\years} in phase 2 to be consistent with non-shocked regions of gas outside the shocked knots. This may only be true for the B2 region, as previous work (Viti et al. 2011) has indicated that a pre-existing, non-homogeneous clump is required for the extant chemistry at B1 to occur. To date, there is no such evidence observed towards B2, hence the homogeneous pre-shock gas assumption. Using this, we can also compute enhancement factors relative to the fractional abundance at $t \approx 0$ \si{\years}, thus allowing direct comparison to the abundances and enhancement factors listed in Table \ref{tab:enhance}.
    
    The chemical network used to compute the abundances considered is based on the network described by \citet{UCLCHEM}. To summarise in brief, we use a reduced form of the UMIST database \citep{umist} to build a network of gas-phase reactions. We also include a dust-grain reaction network that allows for freeze out with hydrogenation and both thermal and non-thermal desorption.

%
%

\section{Results} \label{sec:results}

    \subsection{Model comparison} \label{sec:modelComparison}
    
        \begin{figure}[ht]
            \begin{subfigure}[b]{.5\textwidth}
              \captionsetup{width=.9\textwidth}
              \centering
              \includegraphics[width=.9\linewidth]{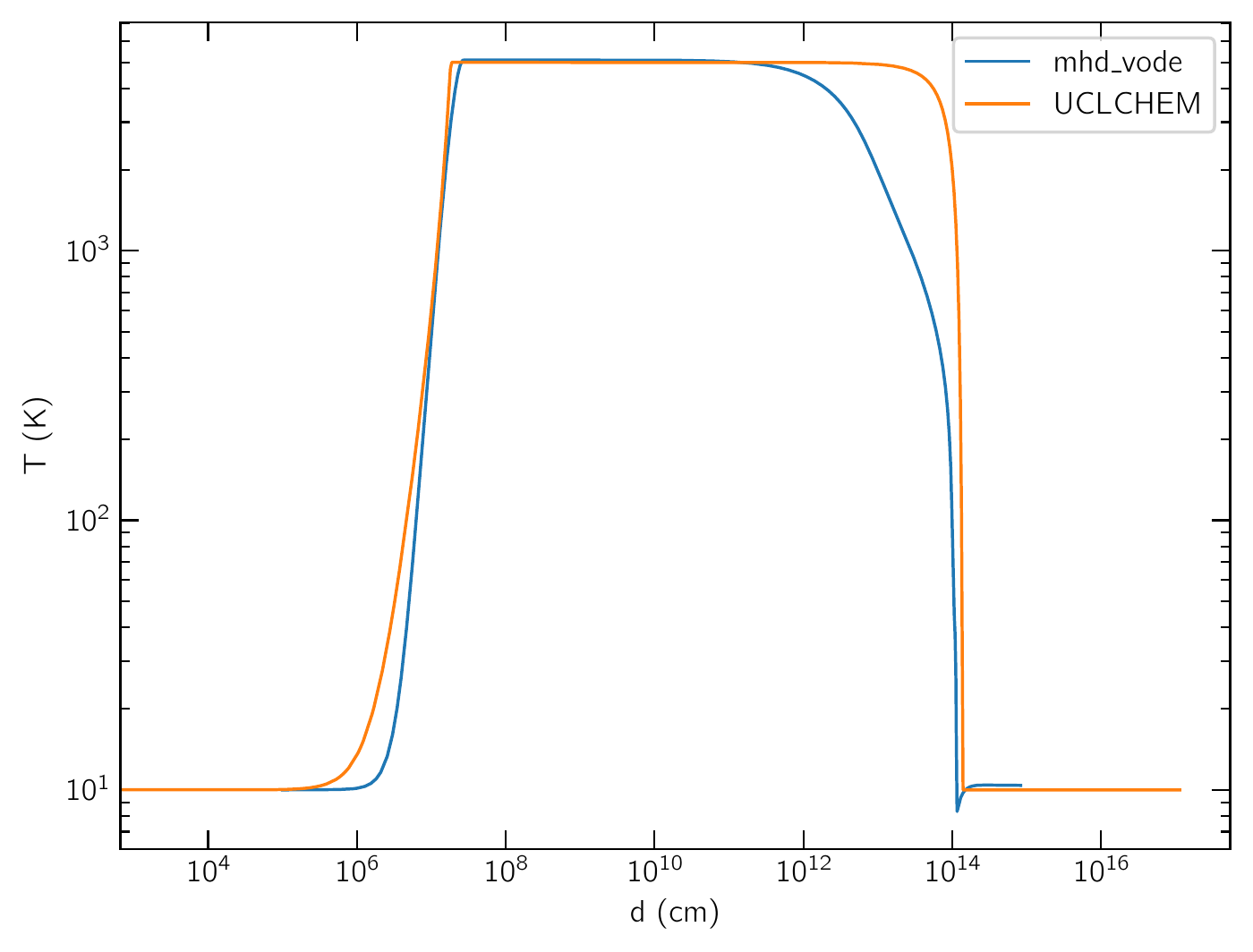}
              \caption{Comparing the temperature profiles for the J-type shock in \lstinline{mhd_vode}, as well as the model presented in this paper.}
              \label{fig:modelT}
            \end{subfigure}
            
            \begin{subfigure}[b]{.5\textwidth}
              \captionsetup{width=.9\textwidth}
              \centering
              \includegraphics[width=.9\linewidth]{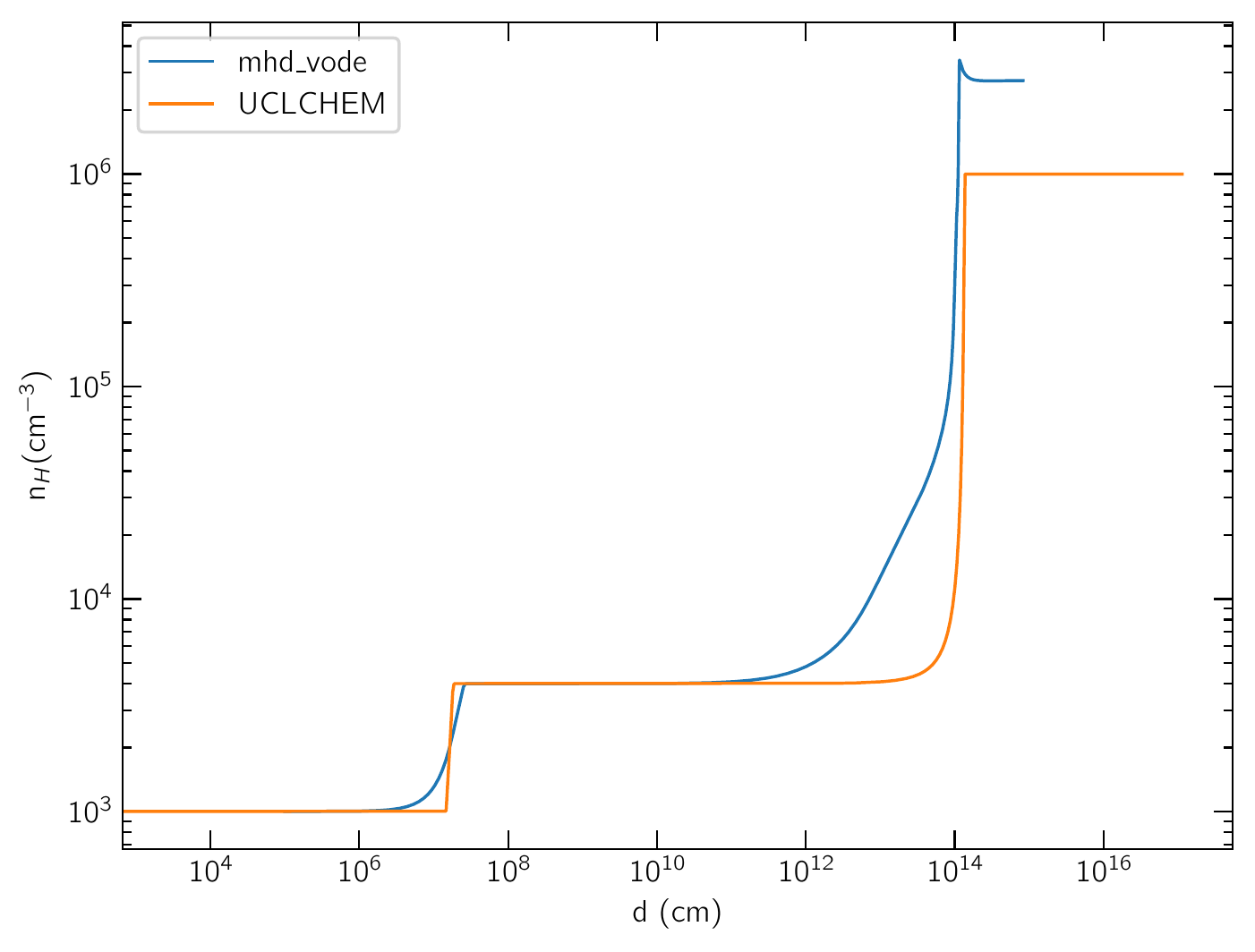}  
              \caption{As Figure \ref{fig:modelT}, comparing the density profiles for the J-type shock in \lstinline{mhd_vode}, as well as the model presented in this paper.}
              \label{fig:modeln}
            \end{subfigure}
            \caption{Comparing the physical structure of a J-type shock with $v=10$ \si{\kilo\meter\per\second} and $n_{H}=10^{3}$ \si{\per\centi\meter\cubed} computed with the model presented in this paper and the \lstinline{mhd_vode} model by \citet{mhd_vode}. Good agreement is observed, despite our approximation not recovering all of the features in the \lstinline{mhd_vode} profile. The model built for \lstinline{UCLCHEM} is also isothermal such that it cools back to its initial temperature, whereas \lstinline{mhd_vode} is not despite it cooling to $\approx 10$ K in this instance.} 
            \label{fig:model}
        \end{figure}
        
        Figures \ref{fig:modelT} and \ref{fig:modeln} show the profiles of temperature $T$ and density $n_{H}$ for both \lstinline{mhd_vode} and the model presented in this work. 
        
        Qualitatively comparing the $T$ profiles in Figure \ref{fig:modelT} we observe good agreement between the \lstinline{mhd_vode} model and the \lstinline{UCLCHEM} model's computation of $T$ in the shock-front described by Equation \ref{eq:shock-front-T}. Both models reach approximately the same $T_{max}$ over an almost identical distance despite the \lstinline{UCLCHEM} model beginning its heating prior to the \lstinline{mhd_vode} model. 
    
        Further agreement is observed until $d \approx 10^{11}$ \si{\centi\meter}, whereby \lstinline{mhd_vode} begins to cool rapidly, further exhibiting an inflexion point at $d \approx 10^{13}$ \si{\centi\meter}, causing $T$ to drop from $5000$ \si{\kelvin} to $300$ \si{\kelvin}. As a result agreement diverges between $10^{11} < d < 10^{14}$ \si{\centi\meter}. This departure is a consequence of \lstinline{mhd_vode}'s radiative cooling, which \lstinline{UCLCHEM} does not implement. 
        
        Furthermore \lstinline{mhd_vode} does not explicitly cool back to its initial temperature, though it does reach an equilibrium temperature very close to that of its initial temperature. Figure \ref{fig:modelT} shows the \lstinline{mhd_vode} model cooling its gas to $\approx 10$ K after $d \approx 10^{14}$ \si{\centi\meter}. The parameterised model presented here explicitly assumes that the gas cools back to $T_{initial}$. In Figure \ref{fig:modelT} this is $10$ \si{\kelvin}. 
        
        Comparisons between $n_{H}$ models in Figure \ref{fig:modeln} show qualitatively less agreement, especially regarding the peak $n_{H}$. However, the \lstinline{UCLCHEM} peak $n_{H}$ is within a factor of $2$ of the \lstinline{mhd_vode} model. 
        
\begin{figure}[ht]
            \centering
            \includegraphics[width=.9\linewidth]{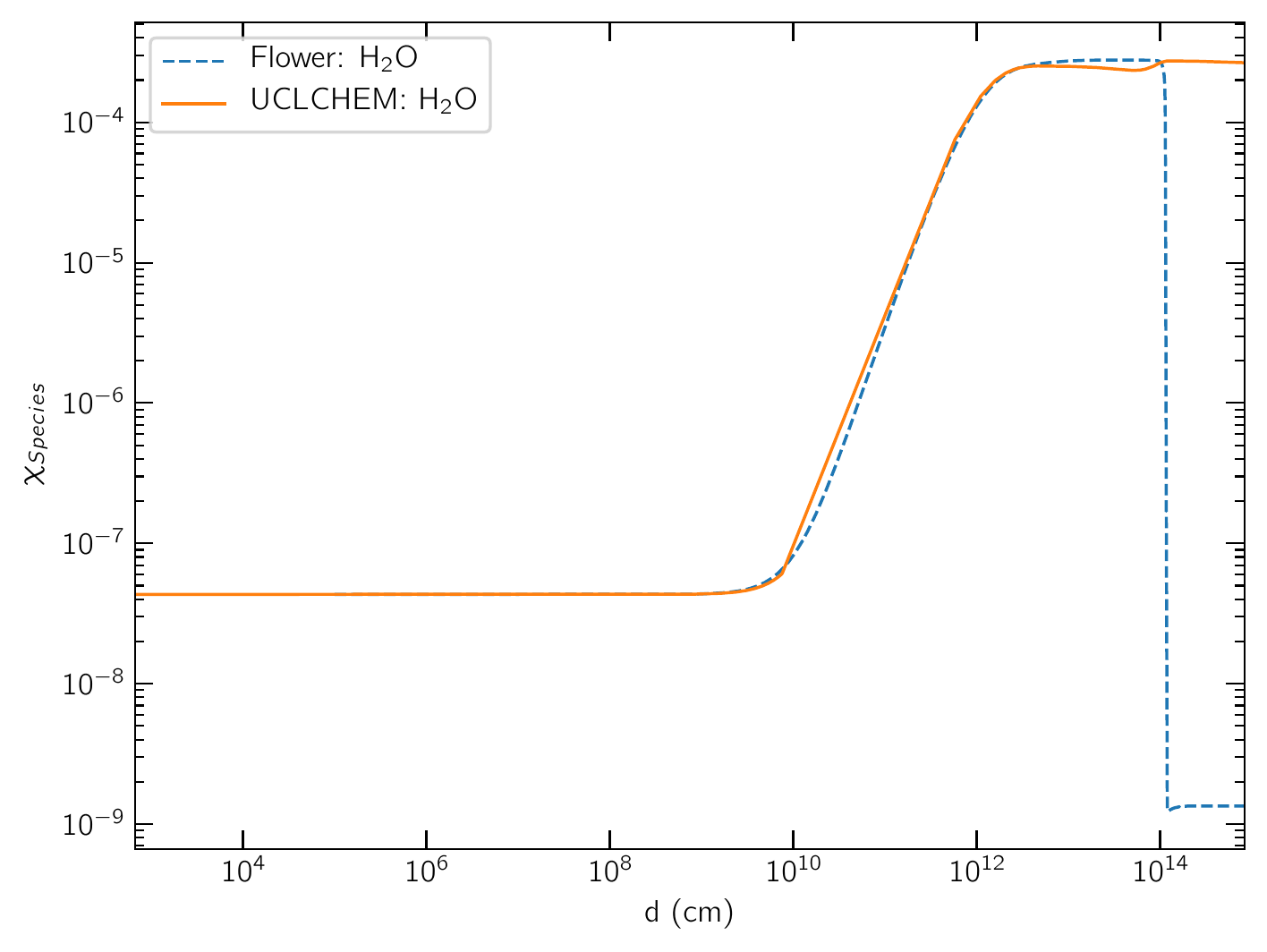}  
            \caption{The evolution of \ce{H_{2}O} during the shock referenced in Figure \ref{fig:model} in both \lstinline{mhd_vode} and \lstinline{UCLCHEM}. Within this figure, sputtering has been deactivated in \lstinline{UCLCHEM} for the purposes of comparison. This implies that only gas-phase chemical reactions are active in these simulations so that the effect of the differences in the temperature profiles between \lstinline{mhd_vode} and our approximation can be fairly evaluated. The abundance evolution of \ce{H_{2}O} up to $d\approx10^{13}$ \si{\centi\meter} is in almost perfect agreement. This is in spite of the lack of inflexion point in both $T$ and $n$ between $10^{11} < d < 10^{14}$ \si{\centi\meter}. This proves that such a departure has negligible effect during the shock. \lstinline{mhd_vode} manually cools \ce{H_{2}O}, hence the decrease in abundance at $d\approx10^{14}$ \si{\centi\meter}. \lstinline{UCLCHEM} does not implement this cooling.}
            \label{fig:h2o-comparison}
        \end{figure}

        The inflexion point highlighted in Figure \ref{fig:modelT} is also present within Figure \ref{fig:modeln} at the same time. Similarly to before, we do not attempt to recover this feature. To assess the effect that this missing feature has on our approximation, and the subsequent chemistry that this model is used to inform, we directly compare the chemistry of \ce{H_{2}O} between \lstinline{mhd_vode} and \lstinline{UCLCHEM}. This is seen in Figure \ref{fig:h2o-comparison}. Importantly, the public version of \lstinline{mhd_vode} used in this study does not include sputtering. Therefore for this comparison, we disable \lstinline{UCLCHEM}'s sputtering treatment to compare chemistry with the same major gas-grain treatments present.
        
        For the same initial conditions, \lstinline{mhd_vode} and \lstinline{UCLCHEM} produce the same \ce{H_{2}O} abundance behaviour despite \lstinline{UCLCHEM} not recovering the observed inflexion point. This is true up to $d=10^{14}$ \si{\centi\meter}, where \lstinline{mhd_vode} radiatively cools \ce{H_{2}O}, causing its abundance to drop sharply. \lstinline{UCLCHEM} does not implement this form of cooling and so the \ce{H_{2}O} abundance does not drop sharply until a much greater distance into the shock.
        
        Given that our model is never more than a factor of 3 away from the \lstinline{mhd_vode} equivalent, and that the shocked \ce{H_{2}O} abundances are in almost perfect agreement, we consider our parameterisation of a J-type shock a good approximation of an equivalent shock model from an ideal-MHD simulation such as \lstinline{mhd_vode}.
        
        Part of our model is the simplifying assumption that the shock is steady-state. This is valid and physically justified as long as the cooling time of the shock is shorter than the time for which the shock velocity and the pre-shock conditions of the gas can change \citep{steadyState}. In our grid runs, we switch back on grain chemistry and assume that the mantle ices instantaneously evaporate if the temperature of the gas $T > 100$ \si{\kelvin}. This is derived from plots within \citet{waterDesorb}. We also assume that any species that have formed in the solid-state on the dust-grain will co-desorb alongside the mantle ices. 
        
        We note that the instantaneous evaporation of the ices in J-type shocks occurs before sputtering takes place. This is fully justified since this is the expected behaviour from the J-type shock’s rapid heating of gas and dust at the sharp shock front. For C-type shocks, we consider both processes, ice evaporation when $T$ exceeds $100$ \si{\kelvin} and sputtering. Since $T$ is significantly lower in C-type shocks, evaporation is less efficient and so sputtering is more effective at releasing a fractional amount of the ices into the gas phase \citep[see][for details on the fractional sputtering technique implemented in \lstinline{UCLCHEM}]{Jimenez-Serra2008}.
        
        The qualitative agreement noted thus far between \lstinline{mhd_vode} model and our parameterised model validates our steady-state assumption for the initial shock conditions applied here. 
    
    \subsection{Identifying J-type shock behaviour}

        To identify unique J-type shock behaviour, we determine the average abundance across the post-shock region\footnote{For the J-type shocks we define the post-shock region as that found between the shock-front and the end of the cooling region; while for a C-type shock, the post-shock region coincides with the length of the dissipation region of the shock.} arising as a result of both J-type and C-type shocks for each model within our grid, and express the ratio of these two average abundances, $\chi(J)/ \chi(C)$. J-type shock enhanced molecules are therefore molecules that have $\chi(J)/ \chi(C) >> 1$. 
        
        To assess the distribution of ratios across the entire grid we bin each model by its values of $v_{s}$ and $n_{H}$ and construct a 2D colour plot. The colour within each bin represents the ratio of the average post-shock abundances, $\chi(J)/ \chi(C)$, up to the dissipation length (or equivalent) for both shock types. 
        
        We also use the enhancement factor, $f_{enhance}$, as a diagnostic. We define $f_{enhance}$ in Equation \ref{eq:enhance}.
        
        \begin{equation}
            f_{enhance} = \frac{\chi(R)}{\chi(0)}
            \label{eq:enhance}
        \end{equation} 

        $\chi(R)$ is the fractional abundance of the shocked molecule, whilst $\chi(0)$ is the fractional abundance of the molecule in a quiescent state. Within this study, we take $\chi(0)$ to be the abundance at simulation time $t \approx 0$ \si{\years} before any sputtering takes place. $f_{enhance}$ is therefore directly comparable to $f$ in Table \ref{tab:enhance}.
        
        This analysis was performed for a range of different known shock-tracing molecules including \ce{CH_{3}OH}, \ce{H_{2}O}, \ce{SO}, \ce{SO_{2}} and \ce{HCN}. We also investigated the behaviour of molecules such as \ce{SiO}, however our analysis indicated that its behaviour was not noteworthy at the considered conditions. We attribute this to our shock velocities $v_{s}$ being too slow to efficiently sputter and form \ce{SiO}.
    
    \subsubsection{\ce{CH_{3}OH}} \label{sec:ch3oh}

        \begin{figure}[ht]
              \centering
              \includegraphics[width=.9\linewidth]{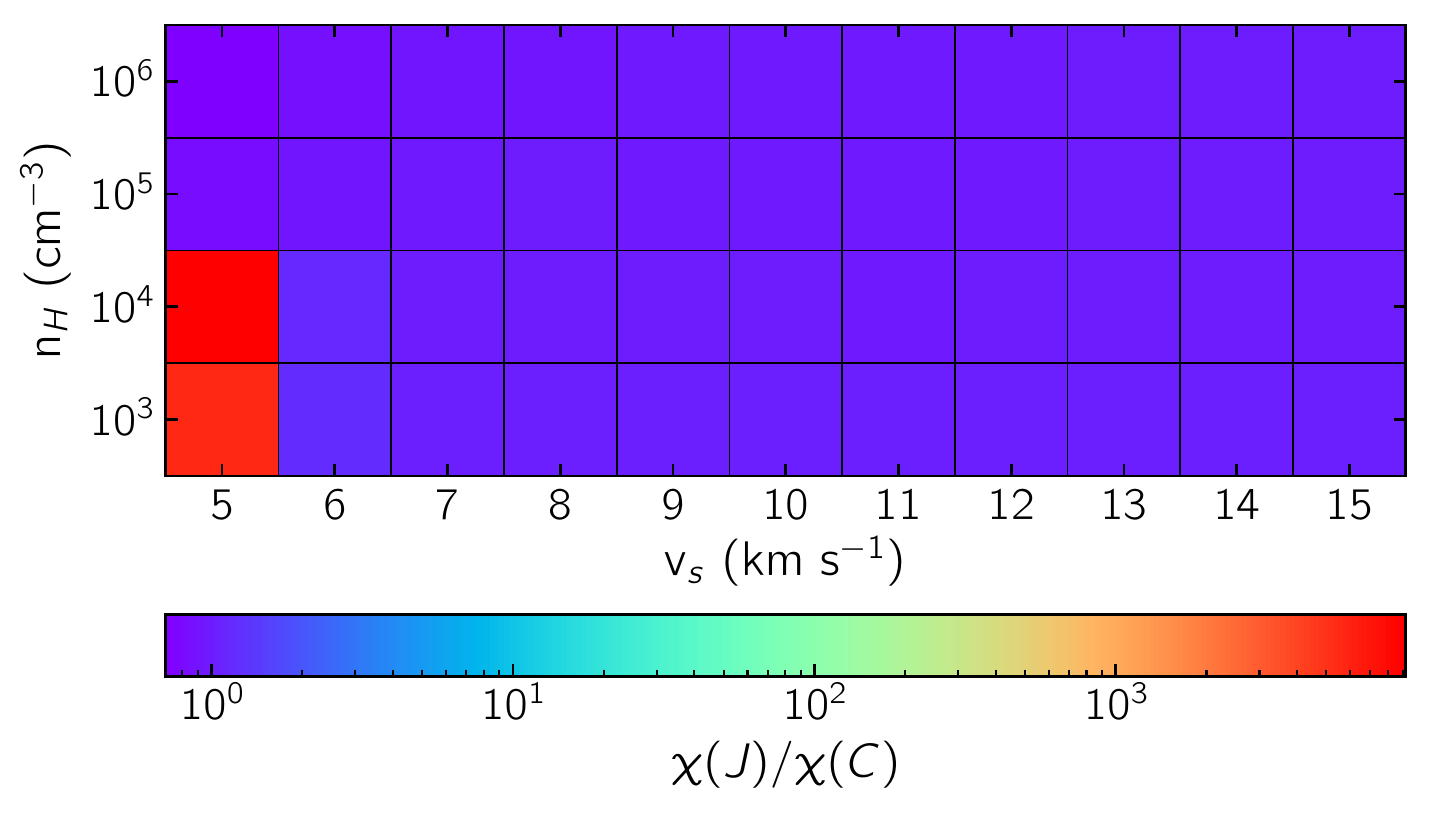}
            
            \caption{Ratio of the average J-type enhanced \ce{CH_{3}OH} abundance to the average C-type enhanced \ce{CH_{3}OH} abundance. As is clear, there is no chemical difference between J-type and C-type enhanced \ce{CH_{3}OH}, except at low $v_{s}$ and low $n_{H}$. This major difference - a factor of $8000$ - arises as a result of the C-type shock failing to sputter grain surface material whilst the J-type shock instantaneously evaporates grain surface \ce{CH_{3}OH}. \label{fig:ch3oh-enhance-ratio}}
        \end{figure}

        \begin{figure}[ht]
            \centering
            \begin{subfigure}[b]{.5\textwidth}
              \centering
              \includegraphics[width=.9\linewidth]{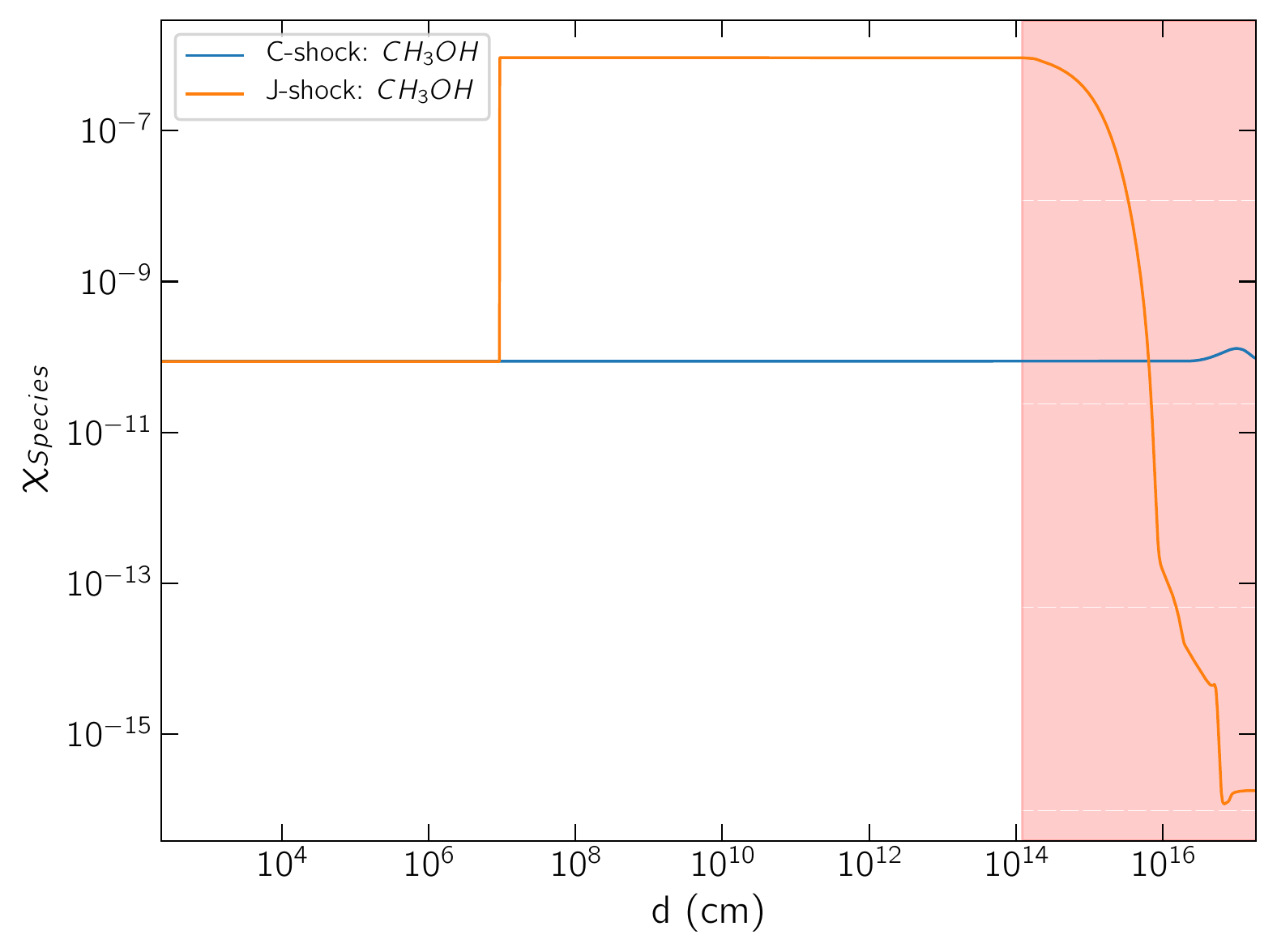}
              \caption{$v_{s} = 5$ \si{\kilo\meter\per\second} and $n_{H} = 10^{3}$ \si{\per\centi\meter\cubed}.}
              \label{fig:ch3oh-v5-n3}
            \end{subfigure}%
            
            \begin{subfigure}[b]{.5\textwidth}
              \centering
              \includegraphics[width=.9\linewidth]{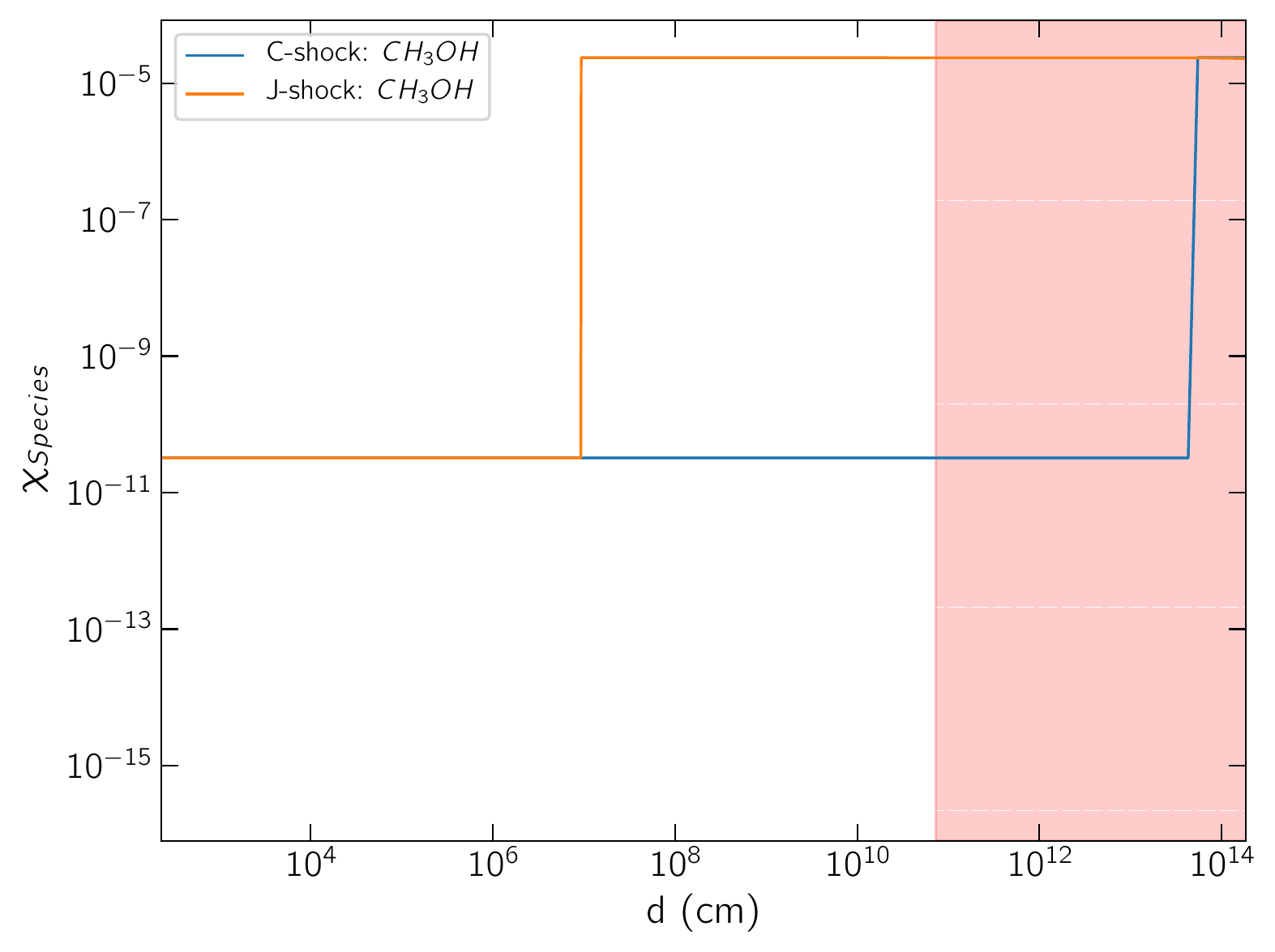}
              \caption{$v_{s} = 5$ \si{\kilo\meter\per\second} and $n_{H} = 10^{6}$ \si{\per\centi\meter\cubed}.}
              \label{fig:ch3oh-v5-n6}
            \end{subfigure}
            
            \caption{\ce{CH_{3}OH} abundances for a shock with initial velocity $v_{s}=5$ \si{\kilo\meter\per\second} and density $n_{H}=10^{3}$ (Figure \ref{fig:ch3oh-v5-n3}) and $n_{H}=10^{6}$ \si{\per\centi\meter\cubed} (Figure \ref{fig:ch3oh-v5-n6}). The shaded red region indicates the region beyond which the J-type shock has cooled to its equilibrium temperature.} 
            \label{fig:ch3oh-v5}
        \end{figure}
        
        Figure \ref{fig:ch3oh-enhance-ratio} shows the ratio of the average post-shock abundances up to the dissipation length (or equivalent) for each shock type. It is computed for C-type shock and J-type shock enhanced \ce{CH_{3}OH} for each model in the grid described in Table \ref{tab:grid}. Within this figure, $\chi(C)$ represents the average gas-phase abundance in a C-type shock achieved up to the dissipation length, whilst $\chi(J)$ is the average gas-phase abundance up to the cooling length for a J-type shock.
        
        Figure \ref{fig:ch3oh-enhance-ratio} shows that there is essentially no difference in chemistry between shock type for \ce{CH_{3}OH}, except the models where $v_{s}=5$ \si{\kilo\meter\per\second} and $n_{H}=10^{3}$ \si{\per\centi\meter\cubed} as well as $n_{H}=10^{4}$ \si{\per\centi\meter\cubed}. 
        
        This unique disparity stems from the stark difference in gas-grain behaviour between shock types under these conditions. As Figure \ref{fig:ch3oh-v5-n3} shows, the \ce{CH_{3}OH} abundance sharply increases as a result of instantaneous evaporation at $d \approx 10^{7}$ \si{\centi\meter} in the J-type shock. In the C-type shock, neither evaporation nor sputtering occurs, meaning the \ce{CH_{3}OH} abundance remains relatively consistent throughout the shock.
        
        This is confirmed in Figure \ref{fig:ch3oh-v5}, which shows the \ce{CH_{3}OH} abundance as a function of distance through both C-type and J-type shocks with velocity $v_{s}=5$ \si{\kilo\meter\per\second} and density $n_{H}=10^{3}$ \si{\per\centi\meter\cubed} and $n_{H}=10^{6}$ \si{\per\centi\meter\cubed}. 
        
        At conditions excluding those already discussed, sputtering becomes efficient, hence the abundance ratios in Figure \ref{fig:ch3oh-enhance-ratio} tending to $1$ uniformly throughout the rest of the grid as a result of \ce{CH_{3}OH} being co-desorbed in a J-type shock and sputtered in a C-type shock in equal measure. Importantly, following injection/sputtering there is minimal subsequent gas-phase chemistry in either shock, hence reinforcing the common abundances achieved in Figure \ref{fig:ch3oh-enhance-ratio} regardless of shock type. 
    
        As a result of the J-type shock's rapid heating, instantaneous evaporation occurs well before any sputtering activity in a C-type shock. In both shocks, the same amount of \ce{CH_{3}OH} is released from the dust-grains owing to self-consistent initial conditions from phase 1 of \ce{UCLCHEM}.

    \subsubsection{\ce{H_{2}O}} \label{sec:h2o}

        \begin{figure}[ht]
            \centering
            \includegraphics[width=.9\linewidth]{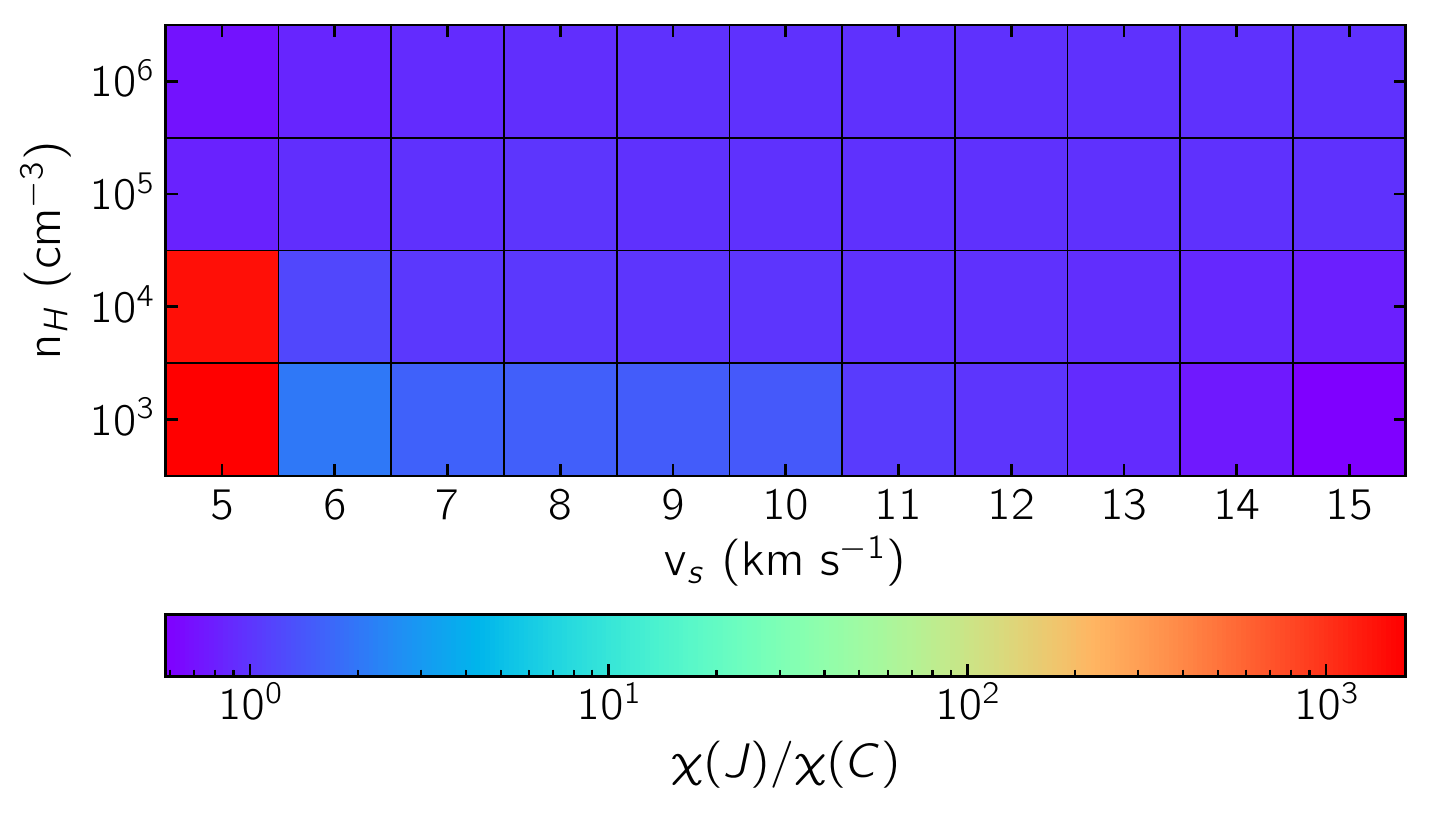}
            
            \caption{Ratio of the average J-type enhanced \ce{H_{2}O} abundance to the average C-type enhanced \ce{H_{2}O} abundance. The largest difference between average shock type abundance is at $v_{s}=5$ \si{\kilo\meter\per\second} and $n_{H}=10^{3}$ \si{\per\centi\meter\cubed} and $n_{H}=10^{4}$ \si{\per\centi\meter\cubed} where the ratio exceeds $1000$.} \label{fig:h2o-enhance-ratio}
        \end{figure}
    
        The abundance ratios for \ce{H_{2}O} is shown in Figure \ref{fig:h2o-enhance-ratio}. Much like \ce{CH_{3}OH} in Section \ref{sec:ch3oh}, \ce{H_{2}O} behaves similarly at $v_{s}=5$ \si{\kilo\meter\per\second} and $n_{H}=10^{3}$ \si{\per\centi\meter\cubed} as well as $n_{H}=10^{4}$ \si{\per\centi\meter\cubed} owing to the same processes; in other words the J-type shock instantaneously evaporates material whilst the C-type shock neither sputters nor evaporates.
        
        Outside of this, the biggest difference between C-type and J-type shocks peaks at $v_{s} < 10$ \si{\kilo\meter\per\second} and $n_{H}=10^{3}$ \si{\per\centi\meter\cubed}. The enhancement factors drop off to $\approx 1$ at velocities and densities greater than these.
        
\begin{figure}[ht]
            \centering
            \begin{subfigure}[b]{.5\textwidth}
              \centering
              \includegraphics[width=.9\linewidth]{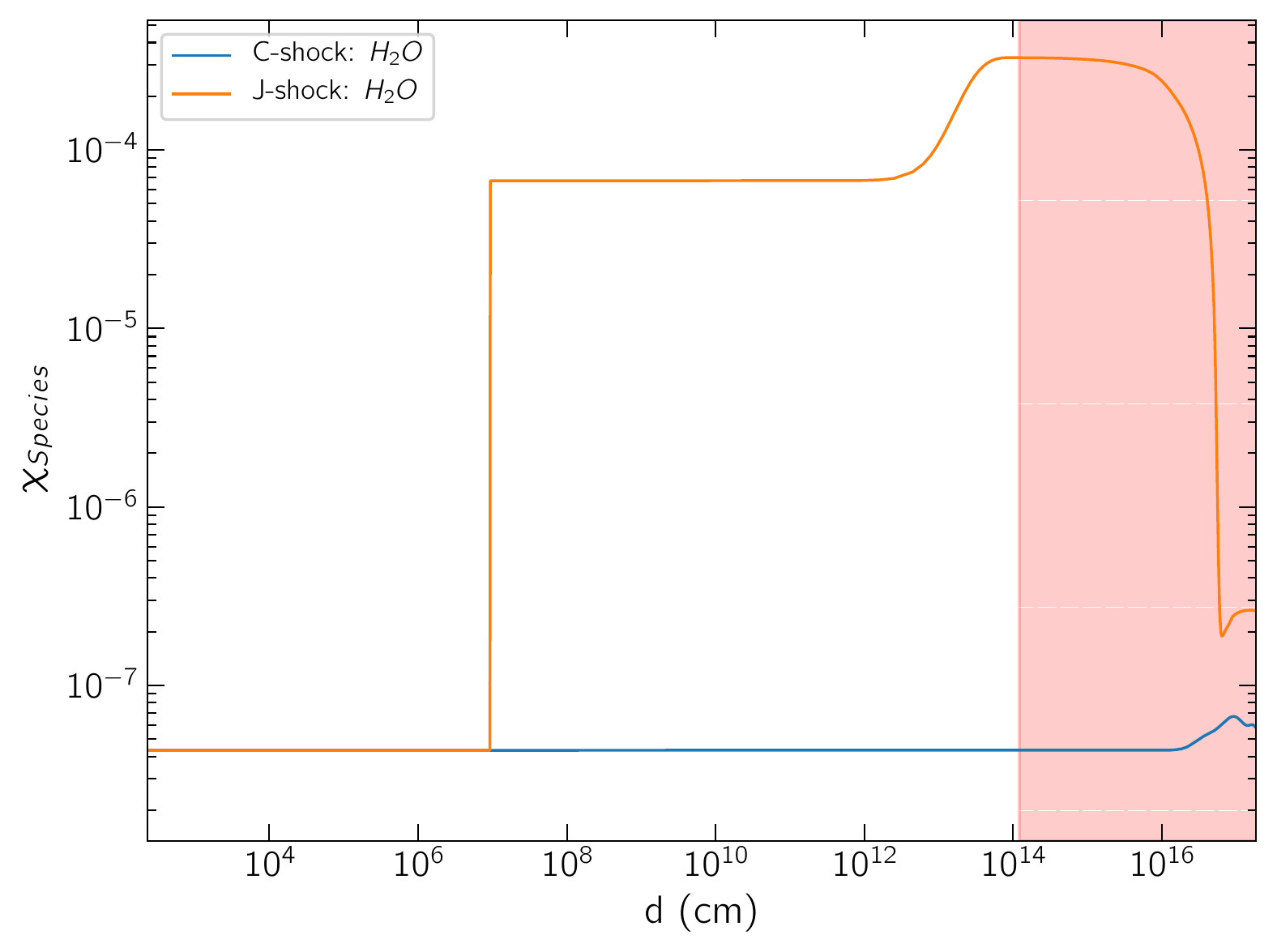}
              \caption{$v_{s} = 5$ \si{\kilo\meter\per\second} and $n_{H} = 10^{3}$ \si{\per\centi\meter\cubed}.}
              \label{fig:h2o-v5-n3}
            \end{subfigure}%
            
            \begin{subfigure}[b]{.5\textwidth}
              \centering
              \includegraphics[width=.9\linewidth]{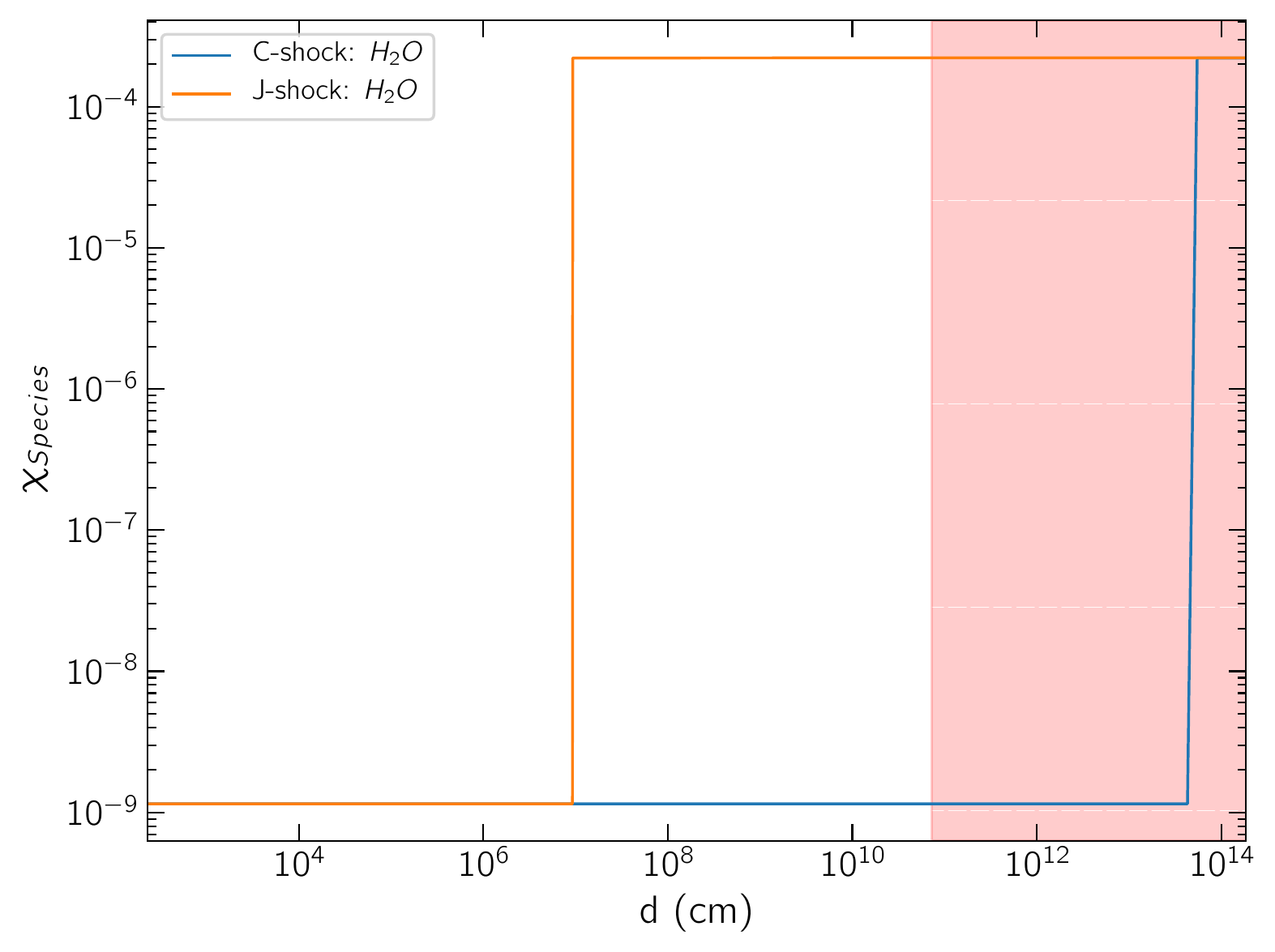}
              \caption{$v_{s} = 5$ \si{\kilo\meter\per\second} and $n_{H} = 10^{6}$ \si{\per\centi\meter\cubed}.}
              \label{fig:h2o-v5-n6}
            \end{subfigure}
            
            \caption{\ce{H_{2}O} abundances for a shock with initial velocity $v_{s}=5$ \si{\kilo\meter\per\second} and density ranging from $n_{H}=10^{3}$ to $n_{H}=10^{6}$ \si{\per\centi\meter\cubed}.} 
            \label{fig:h2o-v5}
        \end{figure}

        Figure \ref{fig:h2o-v5} shows the \ce{H_{2}O} abundances as a function of distance through the shock for C-type and J-type shocks with velocity $v_{s}=5$ \si{\kilo\meter\per\second} and density $n_{H}=10^{3}$ and $n_{H}=10^{6}$ \si{\per\centi\meter\cubed}.
        
        In the J-type shock profiles from Figure \ref{fig:h2o-v5-n3} and Figure \ref{fig:h2o-v5-n6}, the gas phase abundance of \ce{H_{2}O} increases sharply at $\approx 10^{7}$ \si{\centi\meter}. This feature arises as a result of evaporation of the solid state material frozen on to the dust grains, e.g. the ices. The C-type shock may also undergo an increase in gas phase \ce{H_{2}O} at a later time in the shock as a result of sputtering, providing that the initial shock conditions enable the sputtering process. In our models, sputtering does not occur at $v_{s}=5$ \si{\kilo\meter\per\second} and $n_{H}=10^{3}$ \si{\per\centi\meter\cubed} as well as $n_{H}=10^{4}$ \si{\per\centi\meter\cubed}, hence the large difference in average abundance at these models in Figure \ref{fig:h2o-enhance-ratio}.
        
        Post-evaporation features within Figure \ref{fig:h2o-v5} beginto explain the more minor gas-phase enhancement in Figure \ref{fig:h2o-enhance-ratio}. For the J-type shock in Figure \ref{fig:h2o-v5}, the abundance of \ce{H_{2}O} increases to a maximum of $\approx 3\times10^{-4}$, approximately $6$ times the post-evaporation abundance, at around $d \approx 10^{13}$ \si{\centi\meter}. This effect is largest at $n_{H} = 10^{3}$ \si{\per\centi\meter\cubed} and is present as $n_{H}$ increases, though the magnitude of the gas-phase enhancement does decrease as $n_{H}$ increases. At $n_{H}=10^{6}$ \si{\per\centi\meter\cubed} (Figure \ref{fig:h2o-v5-n6}) there is no post-evaporation gas phase abundance change in \ce{H_{2}O}, thus eliminating the effect altogether. 
        
        Investigating the C-type shock in Figure \ref{fig:h2o-v5}, we observe no post-sputtering increase in \ce{H_{2}O}, regardless of $n_{H}$. This, coupled with the decreasing gas-phase enhancement in the J-type as $n_{H}$ increases, results in both shock types tending to the same abundance.
        
        This explains why the largest enhancement is seen at low $v_{s}$, low $n_{H}$. As $n_{H}$ increases, an overall decrease in the post-injection gas phase abundance change is observed, despite the evaporated \ce{H_{2}O} increasing with $n_{H}$. As $v_{s}$ increases, the peak temperature of the shock also rises, allowing gas-phase \ce{H_{2}O} to be destroyed. For a J-type shock, \ce{H_{2}O} destruction begins at $v_{s} = 11$ \si{\kilo\meter\per\second} when $T_{max} > 6000 K$.
        
    \subsubsection{\ce{SO}} \label{sec:so}
 
        \begin{figure}[ht]
            \centering
            \includegraphics[width=.9\linewidth]{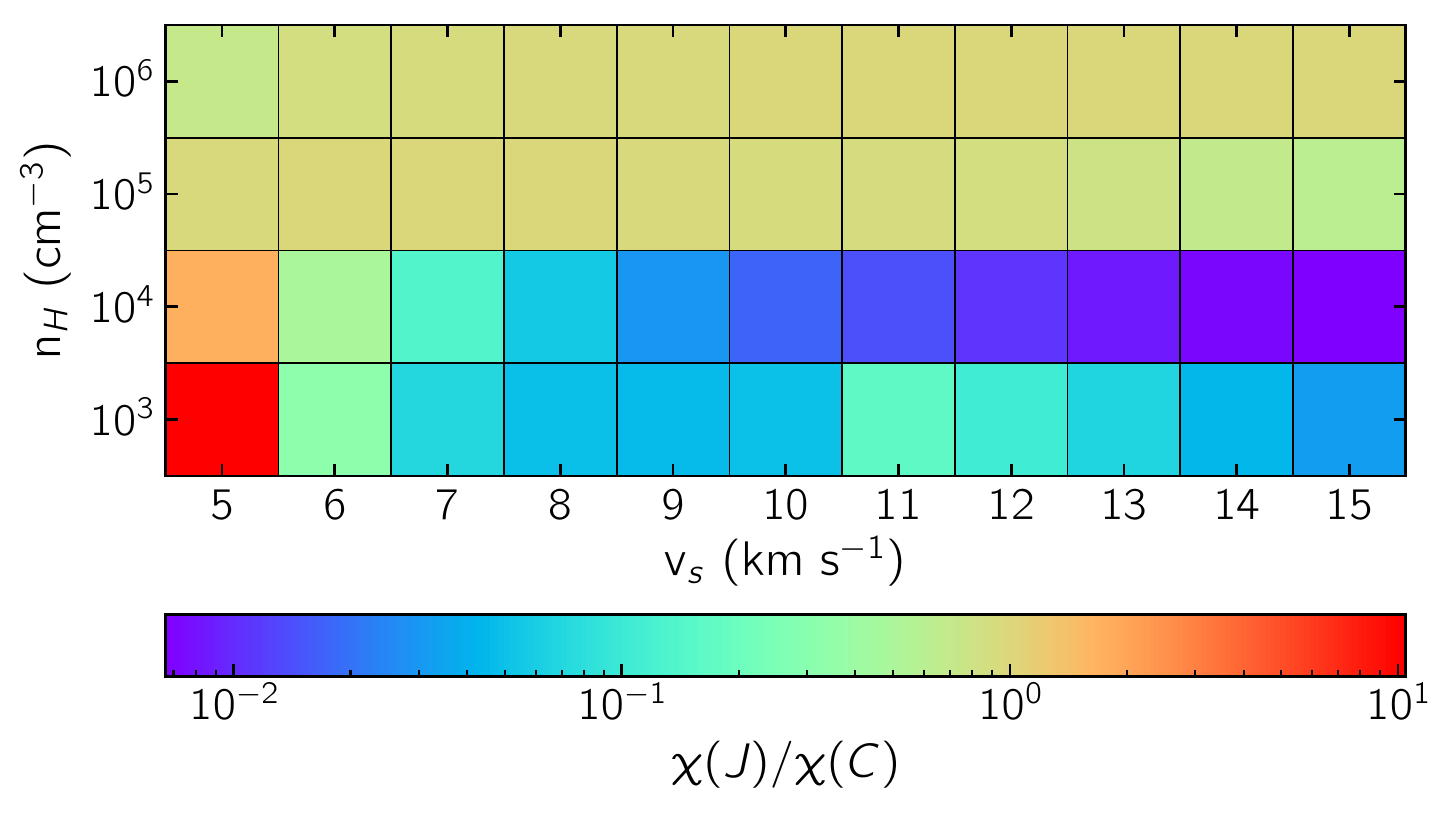}
            
            \caption{Ratio of the average J-type enhanced \ce{SO} abundance to the average C-type enhanced \ce{SO} abundance. The largest difference between peak shock type abundance is at $n_{H}=10^{3}$ \si{\per\centi\meter\cubed}. The shock conditions that produce unique chemistry in this parameter space are those with $n_{H} < 10^{5}$ \si{\per\centi\meter\cubed}.} \label{fig:so-enhance-ratio}
        \end{figure}
        
        Figure \ref{fig:so-enhance-ratio} shows the average abundance ratios for \ce{SO}. Interestingly, Figure \ref{fig:so-enhance-ratio} shows that \ce{SO} is not produced more efficiently in a J-type shock than a C-type shock in our parameter space. In actuality, for $n_{H} > 10^{4}$ \si{\per\centi\meter\cubed} the ratio $\chi(J)/\chi(C) \approx 1$, indicating that at high density both shocks are able to enhance \ce{SO} to similar degrees.
        
\begin{figure}[ht]
            \centering
            \begin{subfigure}[b]{.5\textwidth}
              \centering
              \includegraphics[width=.9\linewidth]{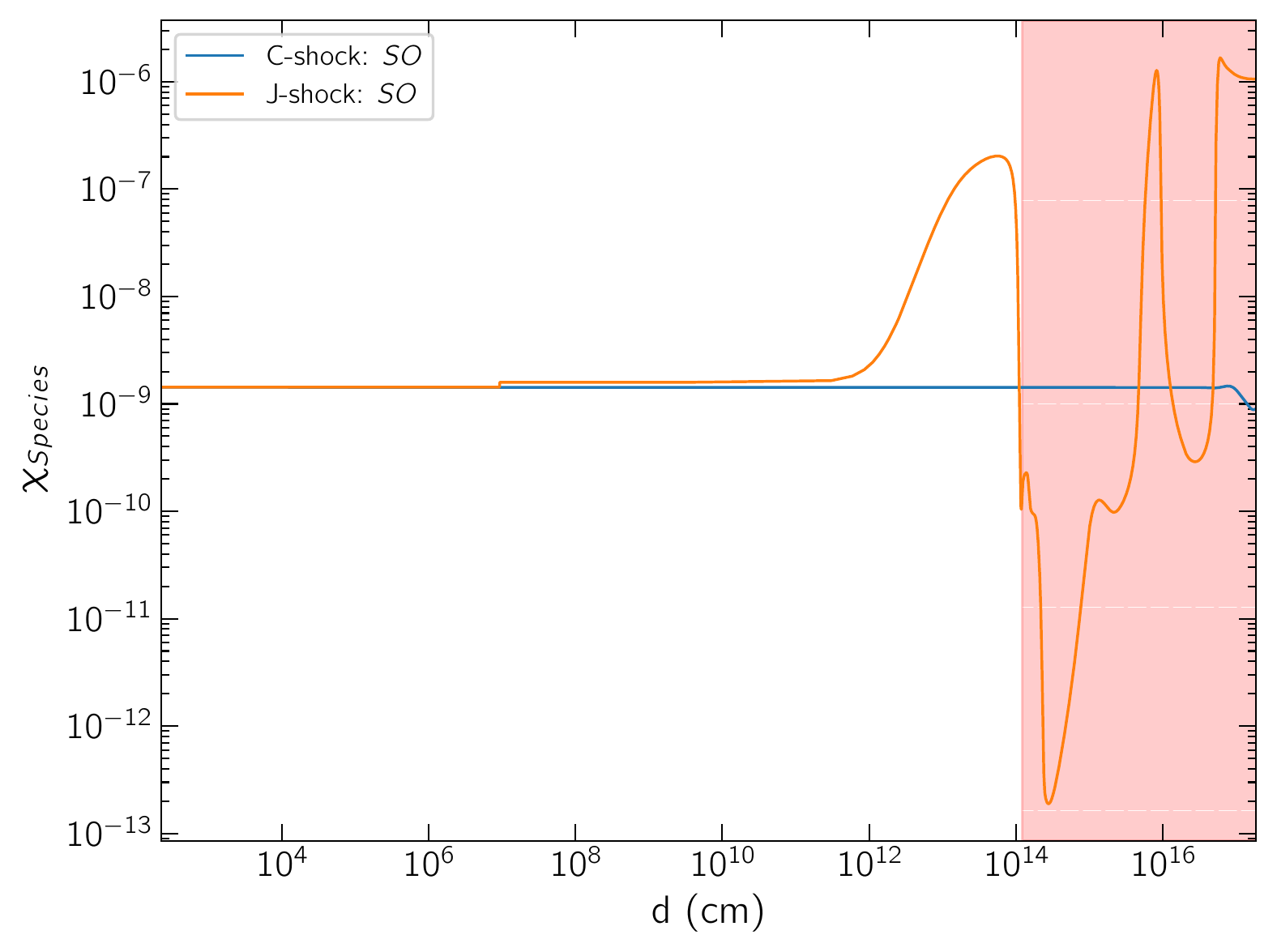}
              \caption{$v_{s} = 5$ \si{\kilo\meter\per\second} and $n_{H} = 10^{3}$ \si{\per\centi\meter\cubed}.}
              \label{fig:so-v5-n3}
            \end{subfigure}%
            
            \begin{subfigure}[b]{.5\textwidth}
              \centering
              \includegraphics[width=.9\linewidth]{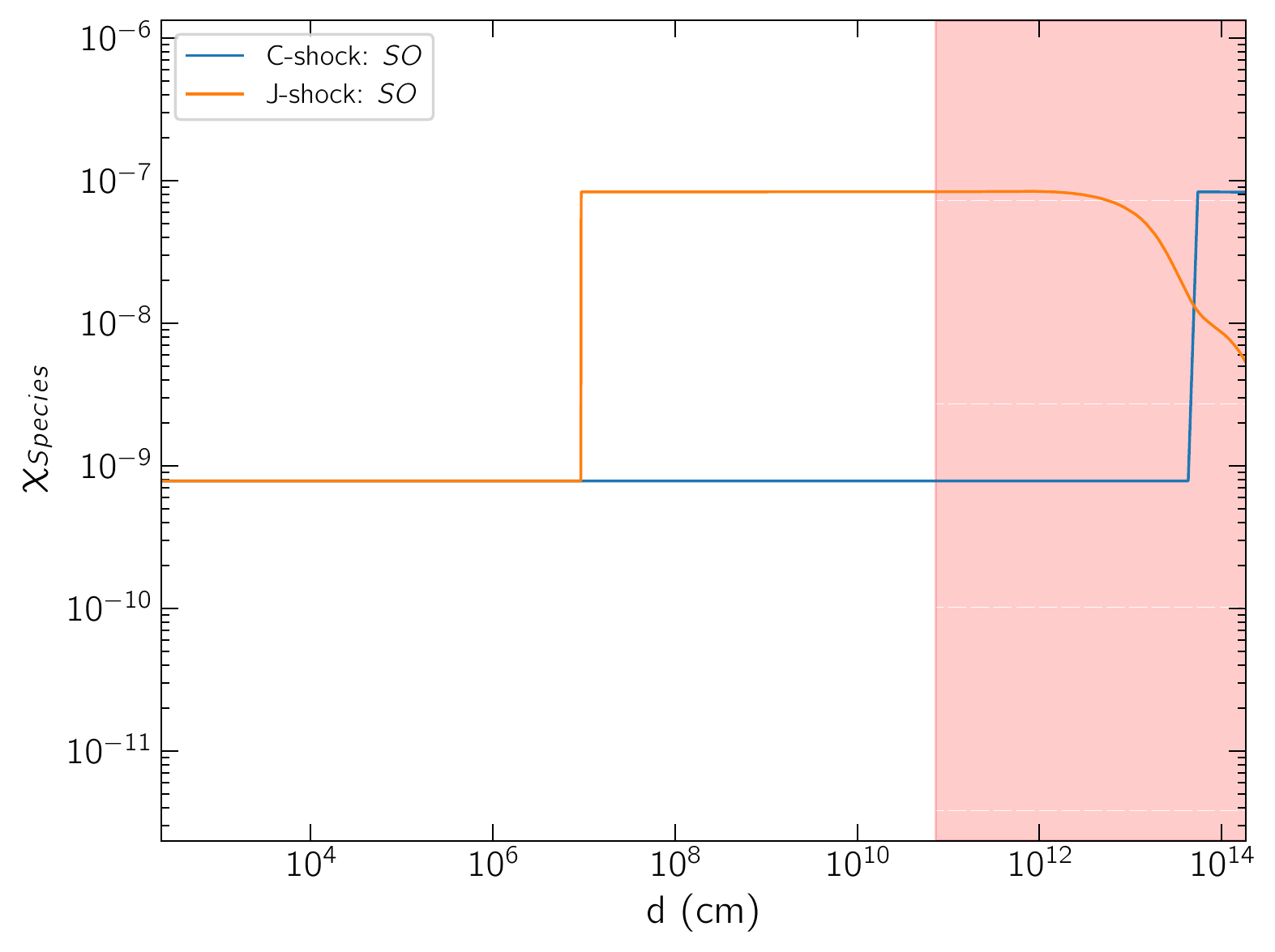}
              \caption{$v_{s} = 5$ \si{\kilo\meter\per\second} and $n_{H} = 10^{6}$ \si{\per\centi\meter\cubed}.}
              \label{fig:so-v5-n6}
            \end{subfigure}
            
            \caption{\ce{SO} abundances for a shock with initial velocity $v_{s}=5$ \si{\kilo\meter\per\second} and density ranging from $n_{H}=10^{3}$ to $n_{H}=10^{6}$ \si{\per\centi\meter\cubed}.} 
            \label{fig:so-v5}
        \end{figure}

        The behaviour of \ce{SO} at $n_{H} < 10^{4}$ \si{\per\centi\meter\cubed} is starkly different. Considering the $n=10^{3}$ \si{\per\centi\meter\cubed} row within Figure \ref{fig:so-enhance-ratio}, it can be observed that the peak ratio of $\approx 10$ occurs at at $v_{s}=5$ \si{\kilo\meter\per\second}. To explain such behaviour, consider the \ce{SO} abundance as a function of distance in Figure \ref{fig:so-v5} for a shock of $v_{s}=5$ \si{\kilo\meter\per\second} with density from $n_{H}=10^{3}$ \si{\per\centi\meter\cubed} and $n_{H}=10^{6}$ \si{\per\centi\meter\cubed}.
        
        Comparing the $v_{s}=5$ \si{\kilo\meter\per\second} and $n_{H}=10^{3}$ \si{\per\centi\meter\cubed} model in Figure \ref{fig:so-enhance-ratio} with the abundance profile for the same initial conditions in Figure \ref{fig:so-v5-n3} begins to explain the peak abundance ratio. It is clear that this arises as a result of the J-type shock injecting \ce{SO} from the grain surface, whilst the C-type shock cannot sputter at these conditions. As $d$ approaches $10^{13}$ \si{\centi\meter} the \ce{SO} abundance peaks at around $10^{-7}$ - an enhancement relative to the initial \ce{SO} abundance of $\approx 100$. However, towards $d \approx 10^{13}$ \si{\centi\meter} the \ce{SO} abundance drops off sharply as \ce{SO} is destroyed. This destruction skews the average \ce{SO} abundance, hence the peak abundance ratio in Figure \ref{fig:so-enhance-ratio} being far smaller than the peak enhancement of $100$. Moreover, $T_{max}$ of a C-type shock of $v_{s}=5$ \si{\kilo\meter\per\second} is $85$ \si{\kelvin}. Such a minimal change in $T$ through the shock is not sufficient to drive any significant gas-phase chemistry, hence the \ce{SO} abundance remaining relatively constant throughout the shock in Figure \ref{fig:so-v5-n3}.
        
        Additionally, as $v_{s}$ increases the C-type shock sputtering becomes more effective whilst the J-type shock destroys \ce{SO} at high $T$, resulting in the average post-shock abundance in a J-type shock being less than the equivalent C-type shock. For example at $v_{s}=15$ \si{\kilo\meter\per\second} and $n_{H}=10^{3}$ \si{\per\centi\meter\cubed}, the J-type shock average abundance is $3\times10^{-2}$ times smaller than the C-type shock equivalent.
        
        This is true of the models at $n_{H}=10^{4}$ \si{\per\centi\meter\cubed} as well, though here we note that the C-type shock sputtering is more efficient therefore exacerbating the differences between average abundance in shock type. Evident here is the J-type shock abundance at $v_{s}=15$ \si{\kilo\meter\per\second} and $n_{H}=10^{4}$ \si{\per\centi\meter\cubed} being $1\times10^{-2}$ times smaller than C-type shock equivalent.

        Figure \ref{fig:so-v5} also shows that as $n_{H}$ increases, the abundances at large $d$ between shock types behaves universally and tends to a similar limit indicating that the dominant destruction mechanism becomes a density limited process. This therefore means that at lower $n_{H}$, the enhancement is governed by a combination of gas-phase and dust-grain chemistry, whilst at large values of $n_{H}$ the enhancement factor is governed by dust-grain chemistry alone.

    \subsection{\ce{SO_{2}}} \label{sec:so2}

        \begin{figure}[ht]
            \centering
            \includegraphics[width=.85\linewidth]{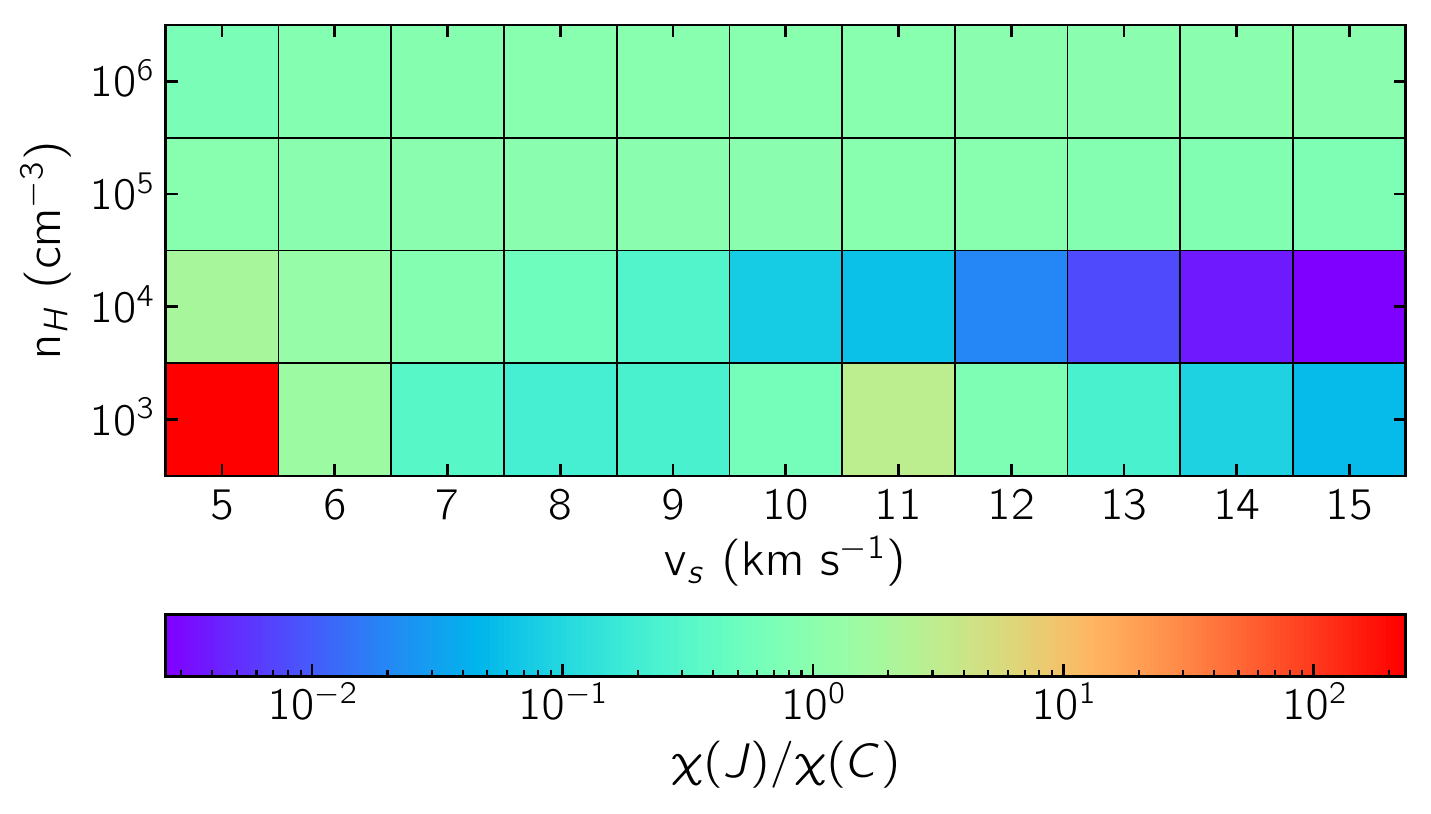}
            
            \caption{Ratio of the maximum J-type enhanced \ce{SO_{2}} abundance to the maximum C-type enhanced \ce{SO_{2}} abundance. The largest difference between peak shock type abundance is at $n_{H}=10^{3}$ \si{\per\centi\meter\cubed} much like the \ce{SO} abundance in Figure \ref{fig:so-enhance-ratio}.} \label{fig:so2-enhance-ratio}
        \end{figure}
        
        Figure \ref{fig:so2-enhance-ratio} shows the abundance ratios for \ce{SO_{2}}. Evident when considering Figure \ref{fig:so2-enhance-ratio} is the similarity between it and the \ce{SO} behaviour in Figure \ref{fig:so-enhance-ratio}. Given that \ce{SO_{2}} can form via \ce{SO} dependent reactions such as \ce{O + SO -> SO_{2}}, the similarity in behaviour is not surprising.
        
        Figure \ref{fig:so2-enhance-ratio} shows largely the same trends as Figure \ref{fig:so-enhance-ratio} did. For instance, we see the same behaviour in $\chi(J)/\chi(C) \approx 1$ at $n_{H}>10^{4}$ \si{\per\centi\meter\cubed} in Figure \ref{fig:so-enhance-ratio}, along with the same model having the same abundance ratio in Figure \ref{fig:so-enhance-ratio}. Curiously, this peak abundance ratio is $\approx 200$, whilst in Figure \ref{fig:so-enhance-ratio} it was $\approx 10$. These global trends and behaviour are expected given the close chemical relationship between \ce{SO} and \ce{SO_{2}}.
        
        \begin{figure}[ht]
            \centering
            \begin{subfigure}[b]{.5\textwidth}
              \centering
              \includegraphics[width=.9\linewidth]{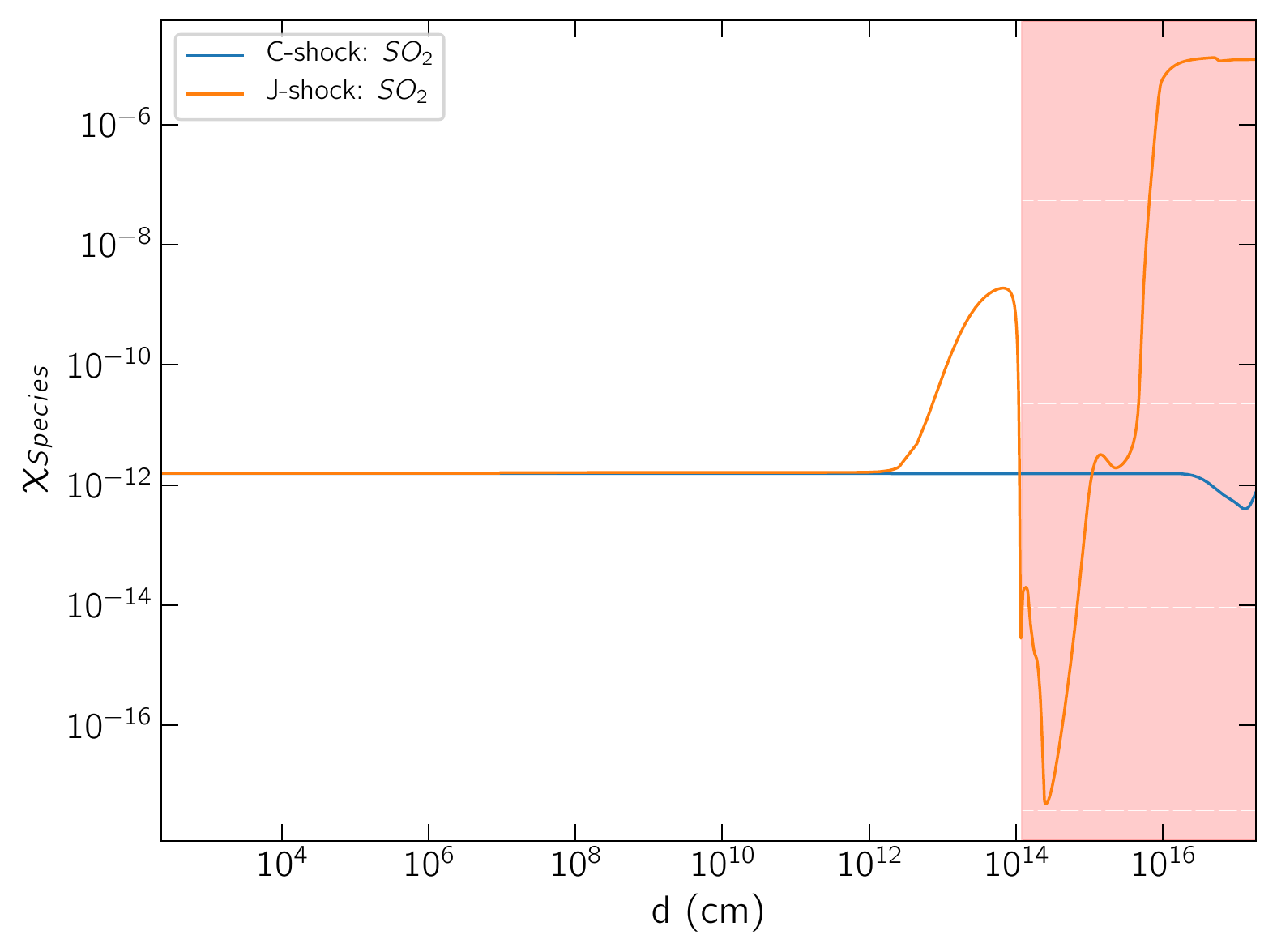}
              \caption{$v_{s} = 5$ \si{\kilo\meter\per\second} and $n_{H} = 10^{3}$ \si{\per\centi\meter\cubed}.}
              \label{fig:so2-v5-n3}
            \end{subfigure}%
            
            \begin{subfigure}[b]{.5\textwidth}
              \centering
              \includegraphics[width=.9\linewidth]{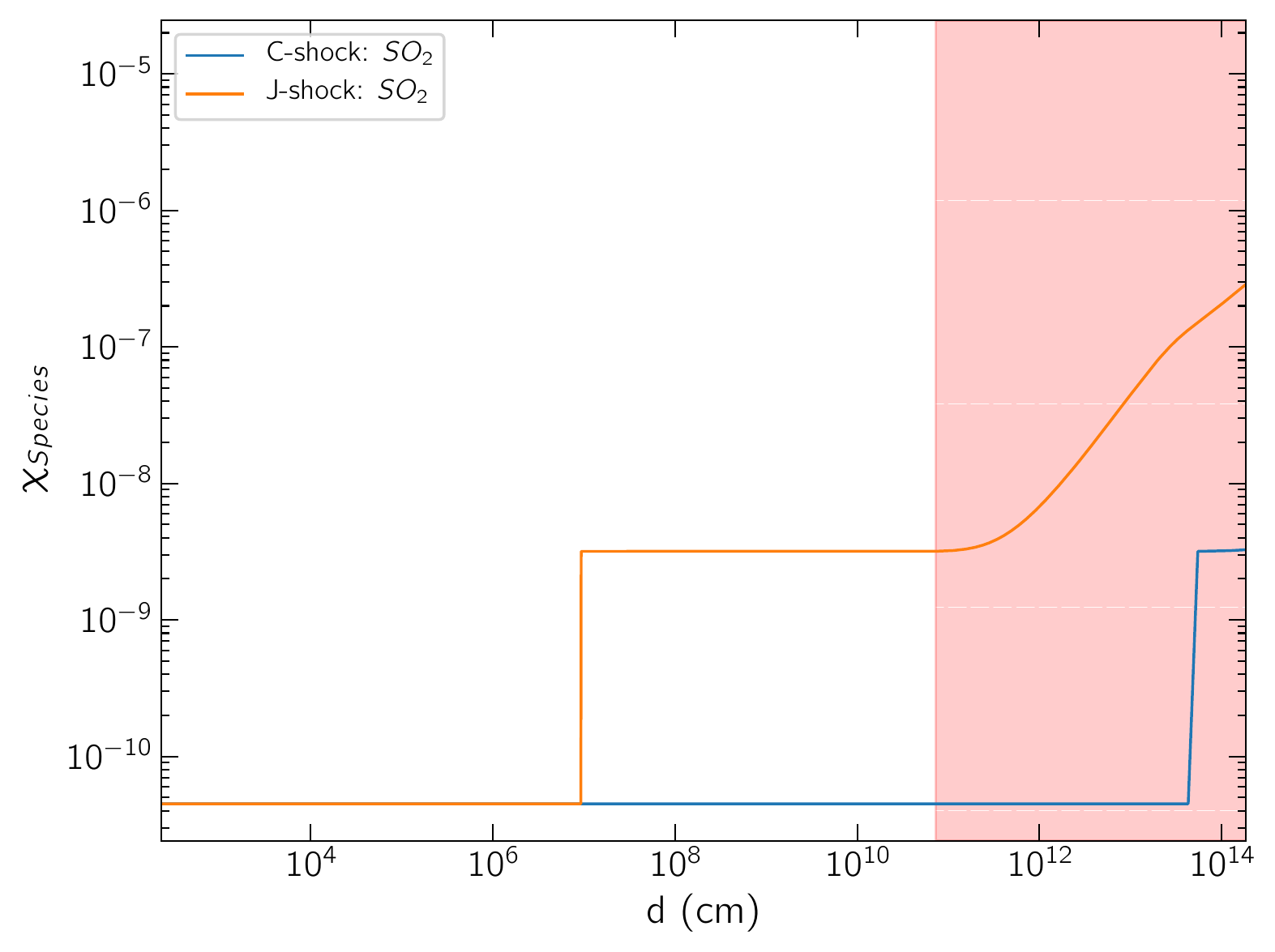}
              \caption{$v_{s} = 5$ \si{\kilo\meter\per\second} and $n_{H} = 10^{6}$ \si{\per\centi\meter\cubed}.}
              \label{fig:so2-v5-n6}
            \end{subfigure}
            
            \caption{\ce{SO_{2}} abundances for a shock with initial velocity $v_{s}=5$ \si{\kilo\meter\per\second} and density ranging from $n_{H}=10^{3}$ to $n_{H}=10^{6}$ \si{\per\centi\meter\cubed}.} 
            \label{fig:so2-v5}
        \end{figure}
        
        Figure \ref{fig:so2-v5} shows the \ce{SO_{2}} abundances as a function of distance for both C-type and J-type shocks at $v=5$ \si{\kilo\meter\per\second} through $n_{H}=10^{3}$ \si{\per\centi\meter\cubed} and $n_{H}=10^{6}$ \si{\per\centi\meter\cubed}. Much like \ce{SO} in Figure \ref{fig:so-v5}, both C-type and J-type shock abundance tend to the same value as $n_{H}$ increases. Furthermore the same behaviour is seen at low $n_{H}$. This implies that any changes to \ce{SO} in a shock should be mirrored - at least in terms of qualitative behaviour - by \ce{SO_{2}} as well.

    \subsection{\ce{HCN}} \label{sec:hcn}

        \begin{figure}[ht]
            \centering
            \includegraphics[width=.85\linewidth]{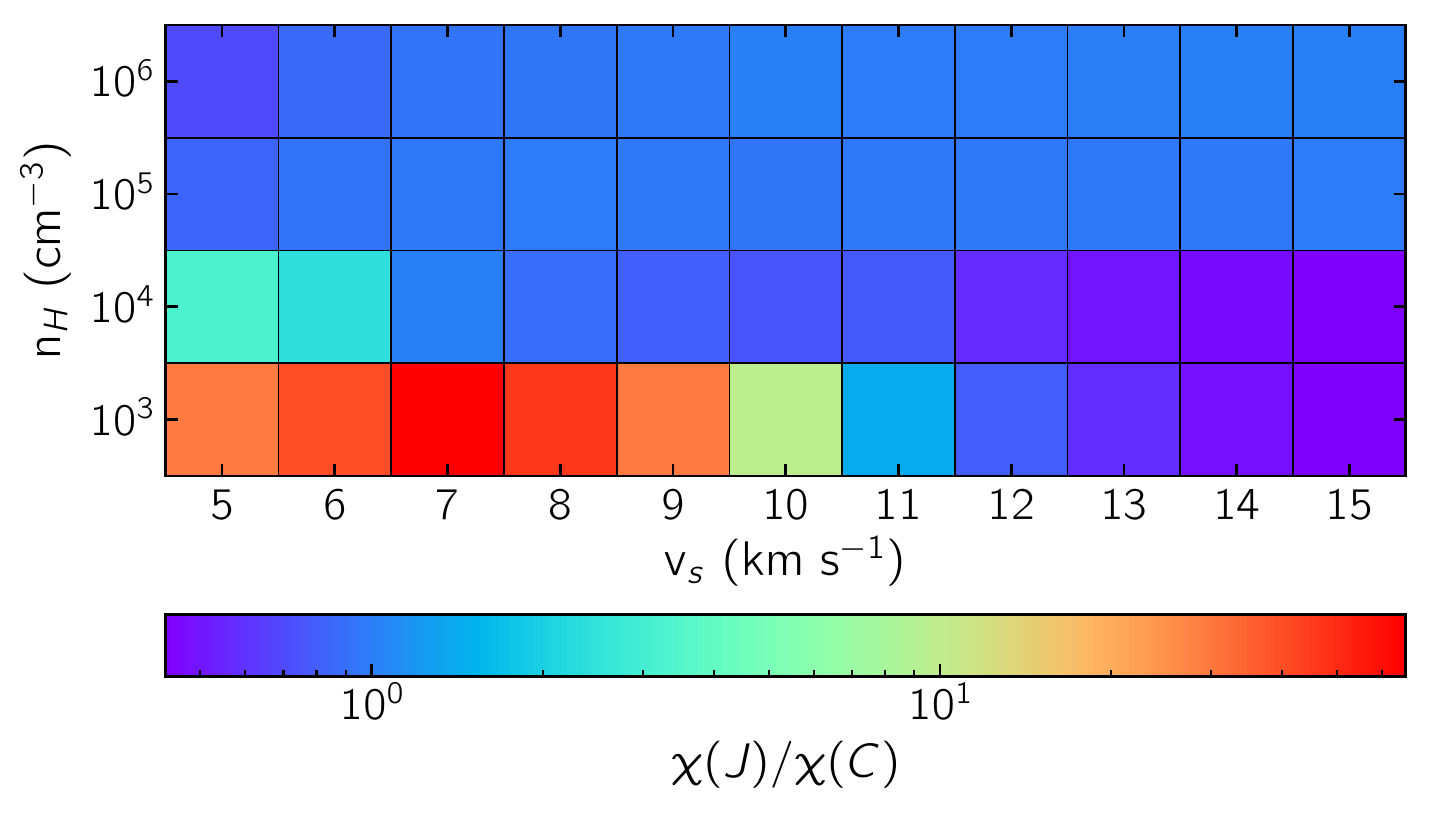}
            
            \caption{Ratio of the average J-type enhanced \ce{HCN} abundance to the average C-type enhanced \ce{HCN} abundance. The largest difference between peak shock type abundance is at $v_{s}<9$ \si{\kilo\meter\per\second} and $n_{H}=10^{3}$ \si{\per\centi\meter\cubed}. High $v_{s}$, low $n_{H}$ shocks show C-type shocks are more efficient enhancers of \ce{HCN} than equivalent J-type shocks.}
            \label{fig:hcn-enhance-ratio}
        \end{figure}
        
        As Figure \ref{fig:hcn-enhance-ratio} shows, the peak abundance ratio occurs at $v_{s} < 9$ \si{\kilo\meter\per\second} and $n_{H} = 10^{3}$ \si{\per\centi\meter\cubed}, with the degree of this ratio decreasing as $v_{s}$ increases. As discussed before in Sections \ref{sec:ch3oh}, \ref{sec:h2o}, \ref{sec:so} and \ref{sec:so2}, it is the stark differences in sputtering and evaporation behaviour between shock types at these conditions that gives rise to this feature.
        
        Much like \ce{SO} and \ce{SO_{2}} beforehand, the ratio for $n_{H} > 10^{4}$ \si{\per\centi\meter\cubed} of Figure \ref{fig:hcn-enhance-ratio} shows very little departure from $1$ indicating that both shock types enhance \ce{HCN} to the same or similar degree. Again similarly to \ce{SO} and \ce{SO_{2}} the enhancements at $v_{s}=12-15$ \si{\kilo\meter\per\second} and $n_{H}=10^{3}-10^{4}$ \si{\per\centi\meter\cubed} indicate C-type shocks are more effective enhancers of \ce{HCN} than a J-type shock. As Table \ref{tab:grid} shows, J-type shocks have far higher $T_{max}$ than an equivalent C-type shock. This implies that between $v_{s}=12-15$ \si{\kilo\meter\per\second} J-type shocks are capable of destroying \ce{HCN} whilst an equivalent C-type shock cannot reach a similarly high $T$, therefore allowing \ce{HCN} to continue formation or not undergo destruction at all. 
        
        
        
        \begin{figure}[ht]
            \centering
            \begin{subfigure}[b]{.5\textwidth}
              \centering
              \includegraphics[width=.9\linewidth]{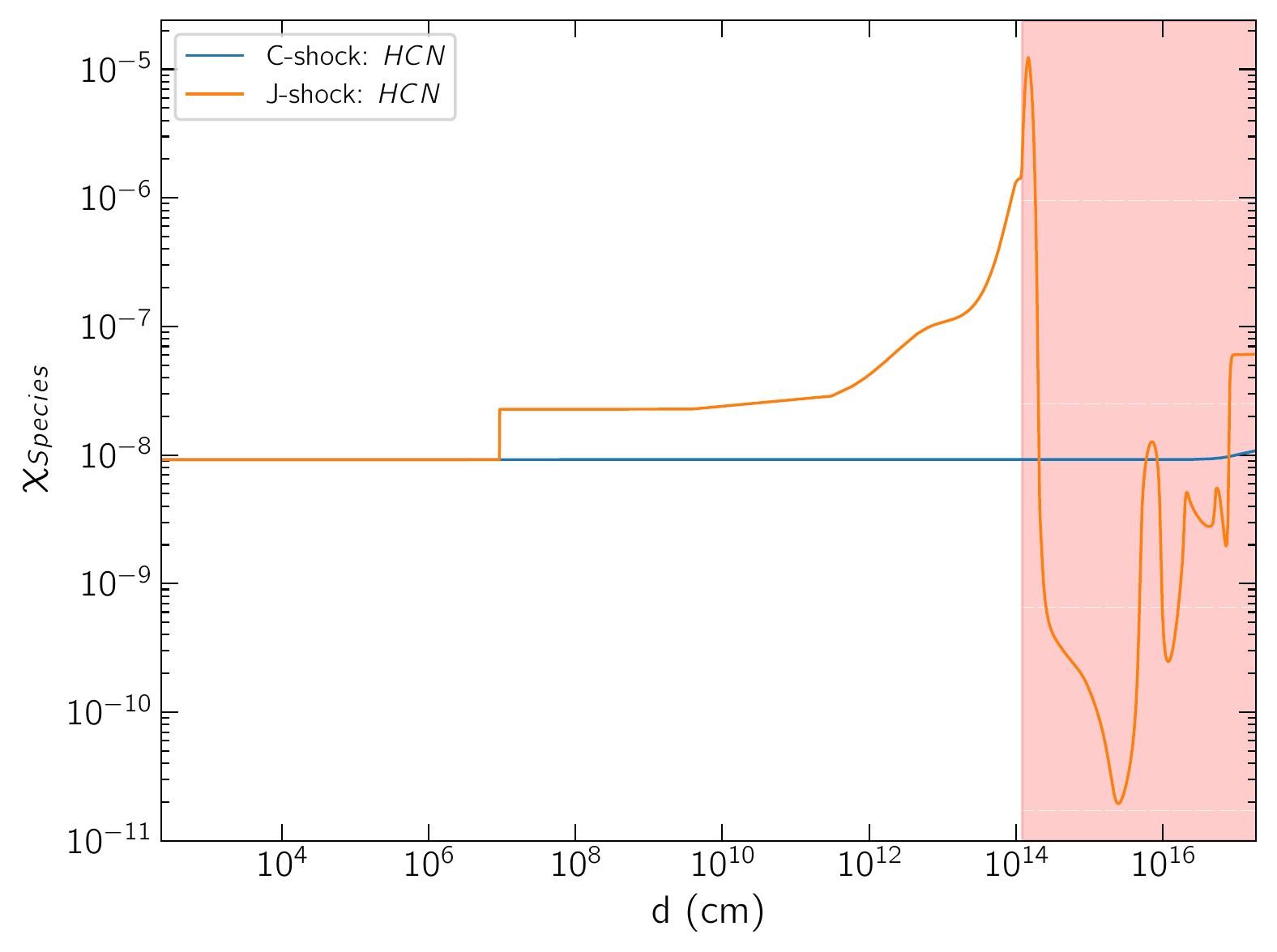}
              \caption{$v_{s} = 5$ \si{\kilo\meter\per\second} and $n_{H} = 10^{3}$ \si{\per\centi\meter\cubed}.}
              \label{fig:hcn-v5-n3}
            \end{subfigure}%
            
            \begin{subfigure}[b]{.5\textwidth}
              \centering
              \includegraphics[width=.9\linewidth]{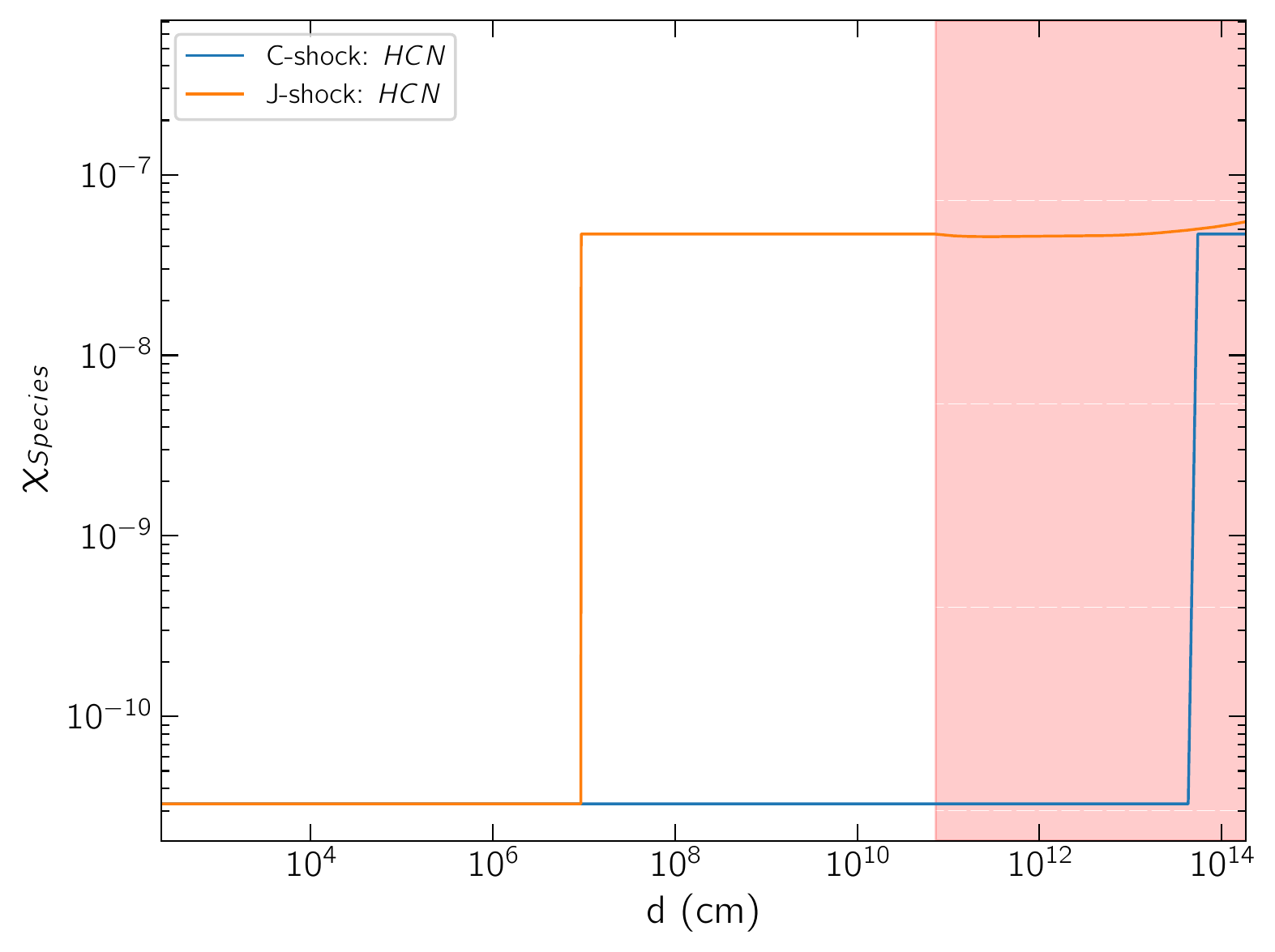}
              \caption{$v_{s} = 5$ \si{\kilo\meter\per\second} and $n_{H} = 10^{6}$ \si{\per\centi\meter\cubed}.}
              \label{fig:hcn-v5-n6}
            \end{subfigure}
            
            \caption{\ce{HCN} abundances for a shock with initial velocity $v_{s}=5$ \si{\kilo\meter\per\second} and density ranging from $n_{H}=10^{3}$ to $n_{H}=10^{6}$ \si{\per\centi\meter\cubed}.} 
            \label{fig:hcn-v5}
        \end{figure}
        
        Individual abundance profiles for \ce{HCN} are shown in Figure \ref{fig:hcn-v5}. As is consistent with other figures, the immediate post-evaporation abundance increases as $n_{H}$. Despite this, the maximal post-shock gas-phase enhancement of \ce{HCN} is at lower density, with the effect dropping off as $n_{H}$ increases. 
        
        Much like previous figures, Figure \ref{fig:hcn-v5-n3} explains why the J-type shock \ce{HCN} abundance is so much greater than the C-type shock \ce{HCN} abundance. Similarly to before, C-type shock sputtering is not possible at $v_{s}=5$ \si{\kilo\meter\per\second} and $n_{H}=10^{3}$ \si{\per\centi\meter\cubed} whilst the J-type shock is capable of instantaneously evaporating the grain-mantle material. Unlike previous molecules however, this behaviour continues up to $v_{s}=12$ \si{\kilo\meter\per\second}. As $n_{H}$ increases to $n_{H}=10^{6}$ \si{\per\centi\meter\cubed} sputtering becomes more efficient and the post-evaporation abundance increases no longer occur. Both of these factors combined allows the \ce{HCN} abundance in both shock types to tend to the same limit of $\approx 5 \times 10^{-8}$. As shown by Figure \ref{fig:hcn-enhance-ratio}, this behaviour occurs at all values of $v_{s}$ for $n_{H}=10^{5}$ \si{\per\centi\meter\cubed} and $n_{H}=10^{6}$ \si{\per\centi\meter\cubed}.

\section{The shocks in L1157-B2} \label{L1157Apply}

     \citet{Vasta2012} observed \ce{H_{2}O} lines towards the B1 and B2 knots of L1157. In conjunction with theoretical shock models, they theorise that J-type shocks could be a prominent source of this emission. Consequently, having thus far found several unique J-type shock chemical distinctions, specifically with respect to \ce{H_{2}O} and \ce{HCN}, we qualitatively apply the results from our grid of models to the B2 region of L1157 in an effort to further categorise the type of shock responsible for its emission. We also compare the results to the measured abundances and enhancement factors in Table \ref{tab:enhance} to further constrain the shock type. Crucially, as mentioned in Section \ref{sec:B2}, the measured abundances are likely subject to large uncertainties owing to the optically thin and thermalised line assumptions required to determine them. 
     
     We focus on B2 and not B1 for a number of reasons. Firstly, \citet{gusdorfSi} theorised that B1 is the result of a combination of C-type and J-type shocks, especially in regards to the \ce{SiO} and \ce{H_{2}} observations. This implies that B2 is also likely to be related to J-type shocks in some form. Further studies such as those by \citet{Vasta2012} also conclude that B2 likely hosts a J-type shock, either singularly or in combination with a C-type shock component. Lastly, given the low and high angular resolution observations of the B2 region by \citet{milenaB2, milenaB213}, it seems that B2 is much more homogeneous than B1. This homogeneity removes any influence of successive shock driven chemistry, making B2 the ideal laboratory with which to test this type of shock diagnostic methodology. 
    
    \subsection{\ce{CH_{3}OH}}
        We showed in Section \ref{sec:ch3oh} that \ce{CH_{3}OH} undergoes no enhancement after its initial release from the dust grains into the gas-phase. This therefore implies that $f(B1)$ and $f(B2)$ in Table \ref{tab:enhance} are dependent only upon the sputtered abundance and not the gas phase chemistry \ce{CH_{3}OH} undergoes.
        
        According to \citet{B2Bright} L1157-B2 has a \ce{CH_{3}OH} abundance of $\approx 2.2 \times 10^{-5}$. \ce{CH_{3}OH}'s minimal gas-phase chemistry therefore means that shock enhancing \ce{CH_{3}OH} to this abundance is solely a result of sputtering and/or evaporation, which itself is a density-dependent effect. This implies that shock enhanced \ce{CH_{3}OH} traces the amount of \ce{CH_{3}OH} on the grains and therefore the density of the pre-shocked region, rather than the shock velocity.
        
        According to Table \ref{tab:enhance} L1157-B2 has a \ce{CH_{3}OH} abundance $500$ times larger than the central protostar, L1157-mm, where the $\chi_{CH_{3}OH}=4.5 \times 10^{-8}$. This is consistent with either a C-type of J-type shock impacting a region of pre-shock density $n=10^{3}$ \si{\per\centi\meter\cubed}. This pre-shock density also produces a pre-shock abundance $\chi_{CH_{3}OH} \approx 2 \times 10^{-9}$, approximately consistent with the pre-shock density measured towards L1157-mm. Importantly this is also consistent with the pre-shock density reported by \citet{L1157Profiles} towards L1157-B2. It remains difficult, however, to use \ce{CH_{3}OH} as a tracer of either shock type or shock velocity owing to its consistent gas-phase chemistry under differing physical conditions.

    \subsection{\ce{H_{2}O}}
        We showed in Section \ref{sec:h2o} that \ce{H_{2}O} can trace J-type shocks at $v_{s}<10$ \si{\kilo\meter\per\second} and $n_{H}=10^{3}$ \si{\per\centi\meter\cubed}.
        
        In application to L1157-B2, however, no enhancement ratio was determined by \citet{Vasta2012}. The \ce{H_{2}O} abundance towards L1157-B2 was determined as $1\times10^{-6}$. This abundance is smaller than all of the immediate post-evaporation/post-sputtering abundances that our models show. These models can, however, recover an abundance similar to this for a shock of $v_{s}<10$ \si{\kilo\meter\per\second} and $n_{H}=10^{3}$ \si{\per\centi\meter\cubed}. Matching the exact measured abundance is only achievable during the post-evaporation \ce{H_{2}O} abundance changes. At $v_{s}>10$ \si{\kilo\meter\per\second}, \ce{H_{2}O} is destroyed in the gas-phase allowing the abundance to drop the order of $10^{-6}$, though as the temperature increases beyond that achieved in $v_{s}\approx12$ \si{\kilo\meter\per\second} the abundance falls well below $10^{-6}$.
        
        Importantly, the best matching shock conditions are also consistent with those determined by \citet{L1157Profiles} as $v_{s}\approx10$ \si{\kilo\meter\per\second} and $n_{H}\approx10^{3}$ \si{\per\centi\meter\cubed}. However, as we do not vary the freeze-out efficiency in this study we cannot conclude with certainty whether the observed abundance is solely a result of the shock or a combination of varying freeze-out efficiency and shock action. A lower freeze-out efficiency and slower shock velocity could reproduce a similar abundance to the observed abundance.
    
    \subsection{\ce{HCN}}
        We showed in Section \ref{sec:hcn} that \ce{HCN} can undergo unique J-type shock enhancement at low $v_{s}$ and low $n_{H}$. As $v_{s}$ and $n_{H}$ increase the abundances in each shock type tend to a similar value, implying that \ce{HCN} can trace low $v_{s}$ and low $n_{H}$ J-type shocks only.
        
        \citet{B2Bright} estimate the \ce{HCN} abundance towards L1157-B2 as $5.5 \times 10^{-7}$, undergoing an enhancement by a factor of $\approx 150$ relative to the L1157-mm \ce{HCN} abundance of $3.6 \times 10^{-9}$. Figure \ref{fig:hcn-v5} shows both C-type and J-type shocks are capable of enhancing \ce{HCN} to the same degree at high $n_{H}$. Figure \ref{fig:hcn-v5} also shows that whilst our models do not recover the exact initial \ce{HCN} abundance of $3.6 \times 10^{-9}$ as measured towards L1157-mm, they are capable of re-producing a value of $\approx10^{-9}$ in the range $n_{H}=10^{3}-10^{5}$ \si{\per\centi\meter\cubed}.
        
        Considering the enhancement factor of $150$, Figure \ref{fig:hcn-enhance-ratio} shows that this is only possible in a J-type shock between $v_{s}=6-8$ \si{\kilo\meter\per\second} and $n_{H}=10^{3}$ \si{\per\centi\meter\cubed}, which is approximately consistent with the shock parameters determined by \citet{L1157Profiles}. 
        
    \subsection{\ce{SO}}

        \citet{B2Bright} report that L1157-B2 is more abundant in \ce{SO} than \ce{SO_{2}}. Crucially, the ranges defined for \ce{SO} abundance in L1157-B1 and L1157-B2 by \citeauthor{B2Bright} intersect, likely because of the close chemical relationship between \ce{SO} and \ce{SO_{2}}.
        
        The initial \ce{SO} abundance measured towards L1157-mm is $5.0 \times 10^{-9}$. Much like \ce{HCN}, our models are capable of recovering an initial \ce{SO} abundance of $\approx 10^{-9}$ in the range $n_{H}=10^{3}-10^{5}$ \si{\per\centi\meter\cubed}.
        
        \citet{B2Bright} measure the abundance of \ce{SO} towards L1157-B2 as $2.0-5.0\times10^{-7}$, which yields an enhancement ratio of $60-100$. Our models, as is evident from Figure \ref{fig:so-enhance-ratio}, show that on average a J-type shock is not able to enhance \ce{SO} to the same degree that an equivalent C-type can produce if both shock types can sputter and/or inject ice material into the gas-phase. This, therefore, implies that any unique \ce{SO} enhancement is the result of a C-type shock.
        
    \subsection{\ce{SO_{2}}}

        According to Table \ref{tab:enhance}, L1157-B2 is subject to an \ce{SO_{2}} enhancement of $\approx 20$ with an initial abundance of $3.0\times10^{-8}$. None of the models produced here are capable of reproducing an initial abundance of this order. This indicates that \ce{SO_{2}} has formed more efficiently towards L1157 than our models would indicate. In actuality, Table \ref{tab:enhance} shows \ce{SO_{2}} being initially more abundant than \ce{SO} by almost an order of magnitude.
        
        Furthermore, the relative similarities in global behaviour between \ce{SO} and \ce{SO_{2}} mean the same conclusion applies here, i.e. within the grid of models, C-type shocks are the producers of unique \ce{SO_{2}} behaviour rather than J-type shocks.
    
%
%
 
\section{Discussion}
    It is clear from Section \ref{sec:results} that the \ce{CH_{3}OH}, \ce{H_{2}O} and \ce{HCN} abundances do allude to a shock component within L1157-B2 of $v_{s}=8-11$ \si{\kilo\meter\per\second} impacting a region of pre-shock density of $n_{H}=10^{3}$ \si{\per\centi\meter\cubed}. However, \ce{CH_{3}OH} does not undergo any gas-phase enhancement unique to a specific shock type, rendering it a reliable tracer of pre-shock density in the majority of cases. 
    
    According to Figure \ref{fig:ch3oh-v5}, \ce{CH_{3}OH} undergoes no post-evaporation gas-phase abundance change. Recent evidence \citep[and references therein]{holdshipCH3OH} indicates that contrary to this finding, \ce{CH_{3}OH} is destroyed in highly energetic, high-temperature events such as shocks. The lack of \ce{CH_{3}OH} destruction in our models could indicate that the network used, in this instance UMIST, may be missing some of the dominant high-temperature destruction routes for \ce{CH_{3}OH}. Alternatively, such findings could point to observations capturing a form of progressive erosion of \ce{CH_{3}OH} from the grain surfaces as first proposed by \citet{progressiveErosion}. Nevertheless, there is sufficient evidence to consider the \ce{CH_{3}OH} abundances determined here as upper limits. 
    
    To address the apparent lack of destruction we performed a test whereby we included several combustion literature derived \ce{CH_{3}OH} destruction routes via collisional dissociation with \ce{H} in our network. We selected these reactions as they are thought to be the most efficient mechanism for \ce{CH_{3}OH} destruction in dissociative J-type shocks \citep{suutarinenCH3OH}. Specifically, the reactions included are \ce{CH_{3}OH + H -> CH_{3} + H_{2}O} \citep{thermalDecompCH3OH}, \ce{CH_{3}OH + H -> H_{2} + CH_{2}OH} \citep{methFlames} and \ce{CH_{3}OH + H -> H_{2} + CH_{3}O} \citep{CHORateCoeff}. However, the addition of these reactions produced complete destruction of gas-phase \ce{CH_{3}OH} at high temperature. This may be because the destruction reactions we included have only been measured under combustion conditions and hence are not necessarily accurate for the densities and temperatures of the ISM environments of our study (Balucani, personal communication, 2019). We therefore cannot draw any definitive conclusions regarding \ce{CH_{3}OH} abundance as a tracer of shock type, beyond the upper-limits derived here, until a follow-up study is performed to investigate the prominent reactions responsible for \ce{CH_{3}OH}'s high-temperature destruction. However, the injection behaviour of \ce{CH_{3}OH} provides an excellent tracer of pre-shock density.
    
    \ce{H_{2}O} and \ce{HCN} both exhibit degrees of enhancement that peak at a factor of $60$ relative to a C-type shock at low $v_{s}$ low $n_{H}$. However the behaviour of both \ce{HCN} and \ce{H_{2}O} at larger $n_{H}$ tends to a common trend between both shock types. Such behaviour indicates that using \ce{H_{2}O} and \ce{HCN} as shock type tracers is only valid, and likely only accurate, at lower values of $v_{s}$ and $n_{H}$. 
    
    Importantly we do not deplete the initial abundance of \ce{S}. As \citet{elementalDepletion} highlight, the observed abundance of \ce{S+} in early-star forming regions matches the approximate cosmic abundance. However, as noted by \citet{sulfurDepletion}, the abundance of \ce{S}-bearing species in molecular clouds is reduced significantly, hence the term 'depletion'. Astrochemical models therefore tend to reduce the elemental \ce{S} abundance to $1\%$ of its cosmic abundance in order to ensure that the \ce{S}-bearing molecular inventory is representative of the region studied, for example a molecular cloud. Given the uncertainty surrounding the exact depletion factors, both universally as well as locally, to introduce a depletion factor here would introduce another degree of freedom and another potentially significant source of error that may potentially yield a larger disagreement between the predictions and observations. We therefore fix our initial \ce{S} abundance as solar. 
    
    Studies such as those by \citet{milenaB2} have shown the B1 and B2 knots themselves have sub-structure. These sub-structural features are likely not thermalised with their surroundings, rendering the observed molecular abundances more abundant than one discrete shock event would produce. It is also possible that B2 may host multiple velocity components, meaning that different molecules may trace different components of the shock. However, the upper limits of the \ce{CH_{3}OH} abundance for L1157-B2 derived here are consistent with a shock of pre-shock density $n_{H}=10^{3}$ \si{\per\centi\meter\cubed} which matches the pre-shock density determined by \citet{L1157Profiles}. 
    
    \citet{L1157Profiles} use \ce{NH_{3}} and \ce{H_{2}O} observations to derive their estimates of the shock parameters. Our \ce{H_{2}O} abundance trends show an immediate post-shock enhanced abundance of between $10^{-4}$ and $10^{-5}$, around a factor $10$ to $10^{2}$ higher than \citet{Vasta2012} would indicate. As mentioned previously, some models are capable of achieving an \ce{H_{2}O} abundance approximately consistent with observations providing that the shock is capable of dissociating \ce{H_{2}O}. Consequently, the observations could be tracing this dissociated \ce{H_{2}O} component. Alternatively, our simulations may over-estimate the formation efficiency of \ce{H_{2}O} on the grains, thus allowing more \ce{H_{2}O} to be released from the grains in sputtering or evaporation than is realistic. 
    
    From Table \ref{tab:enhance} the abundance of \ce{H_{2}O} in L1157 is around $2$ orders of magnitude higher in B1 than B2, indicating that gas-phase \ce{H_{2}O} has either been destroyed in L1157-B2 or that it has had sufficient time to freeze on to the surface of the dust. 

    The behaviour of \ce{SO} and \ce{SO_{2}} in Figures \ref{fig:so-enhance-ratio} and \ref{fig:so2-enhance-ratio} show that they both reliably trace low $v_{s}$, low $n_{H}$ C-type shocks. We have thus far established that L1157-B2 bears some signatures of J-type shock chemistry, but \ce{SO} and \ce{SO_{2}} trace predominantly C-type shocks in the parameter space considered. However, the \ce{SO} and \ce{SO_{2}} behaviour, coupled with the \ce{SO} and \ce{SO_{2}} enhancements in L1157-B2, would seemingly imply that L1157-B2 is host to either a C-type shock component as well as a J-type shock component or a singular component being mixed-type in nature. This mixed-type shock could potentially be a J-type shock that is evolving to take on a more C-type shock structure. Both of these scenarios are consistent with the observed trends. However, to confirm either scenario would require a more detailed model of a mixed-type shock, though observations such as those by \citet{milenaB2, milenaB213} are sufficient to continue exploring this question.
    
    Further observational constraints will surely also be provided by SOLIS \citep{solis}. Such data may allow classification of whether B2 hosts any sub-structure, in turn informing even further constraints on theoretical models of shock action.

%
%

\section{Summary}
    We have developed a parameterised model of an isothermal, planar J-type shock wave as a module to the astrochemical code \lstinline{UCLCHEM}. We compute a grid of models across the parameter space $v_{s}=5-15$ \si{\kilo\meter\per\second} and $n_{H}=10^{3}-10^{6}$ \si{\per\centi\meter\cubed} using our J-type shock model as well as the pre-existing C-type module to quantify the different chemical abundance trends in each shock type. We find the following.

   \begin{enumerate}
      \item Our results show that whilst a theoretical distinction in J-type shock chemistry is found in molecules such as \ce{H_{2}O} and \ce{HCN}, it is largely unique to low $v_{s}$, low $n_{H}$ shocks owing to the extreme temperatures J-type shocks are capable of reaching at high values of $v_{s}$. Furthermore, the largest differences in chemistry between shock types arises as a result of the different sputtering and evaporation behaviours between shock types at low $v_{s}$ and low $n_{H}$. 
      \item We find that \ce{CH_{3}OH} is enhanced by shocks and is a reliable probe of the pre-shock gas density, however we find no difference between its gas-phase abundance between shock type. Recent evidence \citep{holdshipCH3OH} indicates that \ce{CH_{3}OH} is destroyed in high $T$ shocks, indicating that chemical networks lack the high $T$ reactions that permit \ce{CH_{3}OH} to be destroyed. Consequently, astrochemical simulations such as the one presented here can only provide upper limits of the shock-enhanced \ce{CH_{3}OH} abundance.
      \item Finally in our application to the L1157-B2 region, we find that fractional abundances are consistent with both C-type and J-type shock emission, potentially indicating the prevalence a mixed-type shock or multiple shock components. Crucially, however, the similarities in abundances at the initial conditions considered here indicate that the dominant factors affecting shock chemistry are more dependent on the initial shock conditions and not the shock type.
   \end{enumerate}

\begin{acknowledgements}
    T.A.J. is funded by an STFC studentship, and thanks the STFC accordingly. I.J.-S. has received partial support from the Spanish FEDER (project number ESP2017-86582-C4-1-R), and State Research Agency (AEI) through project number MDM-2017-0737 Unidad de Excelencia “María de Maeztu”- Centro de Astrobiología (INTA-CSIC). The authors would like to thank the anonymous referee for their very valuable comments that helped improve the manuscript. We also thank D. Williams, R. Garrod and N. Balucani for their discussions and opinions on aspects of this paper. 
\end{acknowledgements}

\bibliographystyle{aa} 
\bibliography{references}

\end{document}